\documentclass[aps,prb,reprint,groupedaddress]{revtex4-1}

\usepackage{CJK}

\usepackage{graphicx}

\usepackage{amssymb}
\usepackage{amsmath}
\usepackage{bm}

\begin{document}
\begin{CJK*}{GB}{}


\title{Spin Hall Effects in Antiferromagnets}

\author{Sverre A. Gulbrandsen}
\email[]{sverre.a.gulbrandsen@ntnu.no}
\affiliation{Center for Quantum Spintronics, Department of Physics, Norwegian University of Science and Technology, NO-7491 Trondheim, Norway}
\author{Camilla Espedal}
\affiliation{Center for Quantum Spintronics, Department of Physics, Norwegian University of Science and Technology, NO-7491 Trondheim, Norway}
\author{Arne Brataas}
\affiliation{Center for Quantum Spintronics, Department of Physics, Norwegian University of Science and Technology, NO-7491 Trondheim, Norway}

\date{\today}

\begin{abstract}
Recent experiments demonstrate that antiferromagnets exhibit the spin Hall effect. We study a tight-binding model of an antiferromagnet on a square lattice with Rashba spin-orbit coupling and disorder. By exact diagonalization of a finite system connected to reservoirs within the Landauer-B$\ddot{\text{u}}$ttiker formalism, we compute the transverse spin Hall current in response to a longitudinal voltage difference. Surprisingly, the spin Hall conductance can be considerably larger in antiferromagnets than in normal metals. We compare our results to the Berry-phase-induced spin Hall effect governed by the intrinsic contribution in the Kubo formula. The Berry-phase-induced intrinsic spin Hall conductivity in bulk systems shows the opposite behavior. The intrinsic spin Hall effect is drastically reduced when the exchange couplings become large.
\end{abstract}

\maketitle

\end{CJK*}


\section{Introduction}\label{section:introduction}

The spin Hall effect (SHE), first predicted by Dyakonov and Perel in 1971 \cite{DYAKONOVperel1971JETP,DYAKONOV1971PhysicsLettersA}, is one of the cornerstone phenomena in spintronics, as it generates spin currents from charge currents \cite{PRL1999:hirsch,sinova2015spin}. The inverse spin Hall effect (ISHE) enables the detection of spin currents in spintronic devices. The SHE and ISHE are crucial for and facilitate the generation and control of spin currents in current-induced magnetization dynamics \cite{PRL2008Ando,LiuPRL2011stfmr,LiuScience2012} and provide a way to measure such spin dynamics \cite{SaitohAPL2006,CzescchkaPRL2011spinpump,SandwegPRL2011}. Non-local measurements of spin transport in insulators use the SHE as an injection method and the ISHE as a detection mechanism \cite{Cornelissen:NatPhys2005,Cornelissen:PRL2018,Lebrun2018}. Normal metals (NMs) \cite{Valenzuela2006Nature,valenzuela2007electrical,KimuraPRL2007,VilaPRL2007} and semiconductors \cite{DYAKONOV1971PhysicsLettersA,KatoPRL2004,kato2004observation,WunderlichPRL2005,BruneNatPhys2010} with spin-orbit coupling (SOC) are widely used for manifestation of the SHE. Studies of the SHE in superconductors \cite{Takahashi_2011SHEsupercond,Wakamura2015natureMat,PRB2017CamillaSeverin,PhysRevApplied2018Jeon,PhysRevB2018Bergeret}, topological insulators \cite{Khang2018NatMatTopologSHE}, and ferromagnets \cite{Chernyshov2009NaturePhys,Kurebayashi2014NatureNano} have also been performed. Recently, the SHE has been observed in antiferromagnets (AFs) \cite{zhangPRL2014spin,mendesPRB2014,Zhange1600759ScienceAdvances2016}. However, only a limited number of studies on the SHE in AFs \cite{Freimuth:prl2010} have been performed, and a broader understanding is needed to realize its potential. 

AFs have attractive features for spintronics \cite{jungwirth2016antiferromagnetic,baltz2016antiferromagnetismRevModPhys}. The net magnetic moment vanishes such that AFs do not produce stray fields. Therefore, the spin configuration is protected against external magnetic field disturbances \cite{marti2015prospect}. Even so, AFs strongly couple to currents and other materials in many ways \cite{ranChengPRL2014pumping,baltzPRB2018,sverrePRB2018spinTransfer,JohansenPRL2018cavity,ErlandsenPhysRevB.100.100503,SqueezedMagAkashPhysRevB.100.174407}. Long-range spin transport has been demonstrated in AF insulators \cite{Lebrun2018}. Ultrafast spin dynamics have been detected via spin-pumping and the ISHE \cite{Li:Nature2020,Priyanka:Science2020}. The terahertz dynamics of AFs can enable high-speed circuits \cite{ranChengPRL2016oscillator,GomonayNaturePhys2018}. Ultrafast current-induced switching via spin-orbit torques \cite{zeleznyPRL2014neelSOTaf,Wadley2016Science,meinertPhysRevApplied2018}, in combination with nonvolatile magnetic states, makes AFs suitable for information-dense memory devices \cite{Olejnik2017NatCommCuMnAsMemorycell,moriyamaPRL2018sotMemory,Bodnar2018NatureComm,OlejSciAdvances2018}.

Several different mechanisms contribute to the SHE. Distinguishing the different contributions is challenging. Historically, it was proposed that the processes that govern the SHE also described a closely related phenomenon, namely the anomalous Hall effect (AHE) \cite{RevModPhysAnomalousHallEffectNagaosa,PRBanomalousHallEffectButtikerstudy}. In the AHE, a longitudinal charge current induces a transverse charge current. The AHE appears in systems with SOC and broken time-reversal symmetry (e.g., ferromagnets). Based on semi-classical models, and borrowing the nomenclature from the AHE, it is common to separate the spin Hall conductivity as $\sigma_{\mathrm{sH}} = \sigma^{\mathrm{int}}_{\mathrm{sH}} + \sigma^{\mathrm{side}}_{\mathrm{sH}} + \sigma^{\mathrm{skew}}_{\mathrm{sH}}$. The three terms denote the intrinsic spin Hall conductivity, the side-jump contribution \cite{BergerPRB1970sidejump}, and the skew scattering contribution \cite{SMIT195839}. The intrinsic spin Hall conductivity $\sigma^{\mathrm{int}}_{\mathrm{sH}}$ stems from the bulk band structure in the absence of scattering, expressed as a Berry-phase term in the Kubo formula \cite{sinova2015spin}. The contributions from side-jump scattering and skew scattering are called extrinsic, as they are induced by impurity scattering. Semi-classical wave-packet dynamics can also describe the different terms in the spin Hall conductivity \cite{PRB1996ChangNiuWavepackets,PRBwavePacketSundaramNiu}. Within the microscopic Kubo-Streda perturbation theory, the intrinsic-, side-jump-, and skew contributions to the spin Hall conductivity correspond to different types of Feynman diagrams \cite{Prb2007SinovaDiagrams}. 

The intrinsic spin Hall conductivity is relatively simple to evaluate even for complex band structures. In materials with strong SOC, the intrinsic contribution sometimes dominates the SHE. Calculations based only on the intrinsic spin Hall conductivity have given quantitative predictions for the SHE in Pt \cite{PRL2008IntrinsicSHinPt} and other transition metals \cite{PRB2008IntrinsicSeveralMetals}. However, both the intrinsic contribution and the side-jump contribution are independent of the mean free path, and distinguishing between the two contributions is not necessarily straightforward. Furthermore, other contributions can be relevant depending on the materials and scattering lifetimes.

In the well-studied Rashba model \cite{BychkovRashbaOriginal}, a two-dimensional electron gas with parabolic bands and linear-in-momentum Rashba spin-orbit coupling (RSOC), the intrinsic spin Hall conductivity attains a universal value \cite{sinovaUniversalPRL2004}. However, vertex corrections exactly cancel this contribution, and the SHE vanishes \cite{inouePRB2004jul,mishchenkoPRL2004nov,olegPRB2005juni,olgaPRB2005jun,krotkovPRB2006}. This cancellation between the intrinsic part and the side-jump contribution is quite accidental within the Rashba model \cite{SINOVA2006214}. 

Another way of determining the SHE is by considering mesoscopic systems as a multi-terminal scattering problem \cite{buttiker:prb1992}. Within numerically exact methods we can include disorder to all orders. Typically, the spin Hall conductance is evaluated by computing the resulting transverse spin currents in leads in response to a longitudinal voltage difference \cite{nikolicPRBaug2005}. Computations on the mesoscopic spin Hall conductance in NMs with RSOC show that the SHE is quite robust to spin-conserving disorder, and the magnitude of the SHE varies for different system parameters \cite{pareekPRB2002may,hankipiPRBdec2004,ShengPRL2005buttiker,nikolicPRBaug2005}.

Our aim is to shed light on the SHE in AFs. A key question is how the exchange interaction affects the magnitude of the SHE. Furthermore, it is of interest to see if calculations of the Berry-phase-induced intrinsic contribution to the spin Hall conductivity captures the main quantitative and qualitative effects of the staggered field. To address these questions, we consider AFs with RSOC on a square lattice. We investigate the contributions to the SHE in AFs by calculating the SHE in two ways. First, we calculate the mesoscopic spin Hall {\it conductance} in a four-terminal system for an AF with RSOC and spin-conserving disorder by using an exact numerical diagonalization. Second, we compare these results to the intrinsic spin Hall {\it conductivity} (the Berry phase contribution), which is calculated from the linear response Kubo formula of a bulk system in the absence of disorder.

The remainder of this work is organized as follows. In Sec. \ref{section:theory}, we present our model for the mesoscopic spin Hall conductance $g_{\mathrm{sH}}$ describing AFs and NMs. In Sec. \ref{section:kubo}, we calculate the intrinsic spin Hall conductivity $\sigma_{\mathrm{sH}}$ from the Kubo formula. In Sec. \ref{section:numerics}, we present numerical results for $g_{\mathrm{sH}}$ in AFs with spin-conserving disorder, and we compare the mesoscopic spin Hall conductance with the Berry-phase-induced intrinsic spin Hall conductivity $\sigma_{\mathrm{sH}}$. Finally, we discuss and summarize our results in Sec. \ref{section:conclusion}.


\section{Mesoscopic Spin Hall Conductance}\label{section:theory}

\begin{figure}[h]
	\centering
	\includegraphics[width=0.75\linewidth]{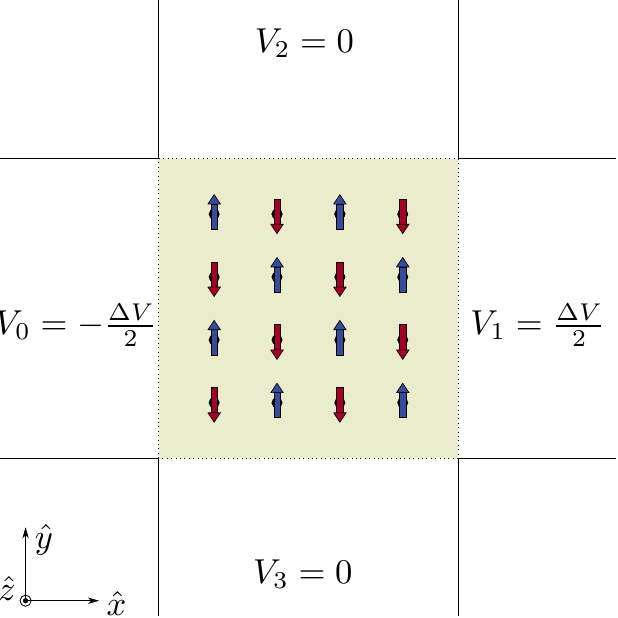}
\caption{Four-terminal scattering problem. The scattering region includes RSOC and antiferromagnetically ordered localized spins. The localized spins of nearest neighbors point in opposite directions, as shown by the blue and red arrows. The scattering region has area $L^{2}$ and is connected to four NM leads. The leads are labeled $l=0,1,2$ and $3$, which correspond to the left, right, top and bottom leads, respectively. Each lead is in contact with an out-of-equilibrium reservoir, where the electric potential is $V_{l}$ for $l=0,1,2$ and $3$. }
	\label{Fig:fourterminalAF}
\end{figure}

We study a four-terminal system within the Landauer-B$\ddot{\text{u}}$ttiker formalism \cite{buttiker:prb1992}. The system includes an AF with RSOC confined to a finite size scattering region connected to four leads. The leads are NMs. Each lead connects to out-of-equilibrium reservoirs. In the reservoirs, the distributions of the electrons are controlled by externally applied voltages $V_{l}$, where $l=0,1,2$ and $3$ labels each terminal (or, equivalently, left, right, top and bottom), as shown in Fig. \ref{Fig:fourterminalAF}. The temperature of each reservoir is zero.
Using the scattering matrix of the system, we can calculate the charge currents $I^{\mathrm{c}}_{l}$ and spin currents $\boldsymbol{I}^{\mathrm{s}}_{l}$ in each lead $l$ driven by the electric potentials in the reservoirs. The spin current is polarized and has three components, $\boldsymbol{I}^{\mathrm{s}}_{l} = (I^{\mathrm{s},\mathrm{x}}_{l}, I^{\mathrm{s},\mathrm{y}}_{l}, I^{\mathrm{s},\mathrm{z}}_{l})$.

We assume that transport is in the linear regime. The electric potentials $V_{l}$ are as follows (see Fig. \ref{Fig:fourterminalAF}). A potential difference $\Delta V = V_{1} - V_{0}$ exists between the right and left reservoirs, with $V_{1} = \Delta V / 2$ and $V_{0} = -\Delta V / 2$. The electric potentials in the top and bottom terminals are set to zero: $V_{2} = 0$ and $V_{3} = 0$. The potential difference $\Delta V$ drives charge currents, which, via SOC, induce transverse spin currents. 

In the Landauer-B$\ddot{\text{u}}$ttiker formalism, the out-of-equilibrium charge currents $I^{\mathrm{c}}_{l}$ from lead $l$ into the scattering region are expressed as
\begin{equation}\label{eq:charge_current_LB_definition}
I^{\mathrm{c}}_{l} = \frac{e^{2}}{2\pi\hbar} \sum_{l'}\left( g_{l'l}V_{l} - g_{ll'}V_{l'}  \right) \, ,
\end{equation}
where $g_{ll'}$ is the dimensionless electrical conductance from lead $l'$ to lead $l$, and the sum is over all leads $l'$. Here, $\hbar$ is the reduced Planck constant, and $-e$ is the electron charge. In linear transport, the conductances are evaluated at the Fermi energy $E_{\mathrm{F}}$. The scattering matrix is unitary, and charge currents are conserved.

To calculate spin currents, one can introduce spin-resolved conductances $g_{ll'}\rightarrow g^{ss'}_{ll'}$. In this notation, $g^{ss'}_{ll'}$ is the conductance for spin $s'$ originating in lead $l'$ transferred to spin $s$ in lead $l$ \cite{nikolicPRBaug2005}. The spin is either $s=\uparrow$ or $s=\downarrow$ along a chosen quantization axis in the corresponding lead. The electrical conductances are related to the spin-resolved conductances as
\begin{equation}
g_{ll'} = g^{\uparrow\uparrow}_{ll'} + g^{\uparrow\downarrow}_{ll'} + g^{\downarrow\uparrow}_{ll'} +g^{\downarrow\downarrow}_{ll'} \, .
\end{equation}
The spin currents in the leads can be expressed in a form similar to Eq.\, \eqref{eq:charge_current_LB_definition} by defining the spin-resolved quantities
\begin{subequations}\label{eqs:spinCondsOutIn}
\begin{align}
g^{\mathrm{out}}_{ll'} &= g^{\uparrow\uparrow}_{ll'} -  g^{\uparrow\downarrow}_{ll'} + g^{\downarrow\uparrow}_{ll'} - g^{\downarrow\downarrow}_{ll'} \, , \\
g^{\mathrm{in}}_{ll'} &= g^{\uparrow\uparrow}_{ll'} +  g^{\uparrow\downarrow}_{ll'} - g^{\downarrow\uparrow}_{ll'} - g^{\downarrow\downarrow}_{ll'} \, .
\end{align}
\end{subequations}
The out-of-equilibrium spin current $I^{\mathrm{s,}\mathrm{z}}_{l}$ in lead $l$ is
\begin{align}
I^{\mathrm{s,}\mathrm{z}}_{l} = \frac{(-e)}{4\pi}\sum_{l' \neq l} \left( g^{\mathrm{out}}_{l'l}V_{l} - g^{\mathrm{in}}_{ll'}V_{l'}    \right) \, ,
\label{eq:spin_current_LB_definition}
\end{align}
with a flow direction towards the scattering region. Similarly, we calculate $I^{\mathrm{s},\mathrm{x}}_{l}$ and $I^{\mathrm{s},\mathrm{y}}_{l}$ by changing the spin quantization axes.

Correspondingly, the spin Hall conductance has three polarization components $g^{\mathrm{x}}_{\mathrm{sH}}$, $g^{\mathrm{y}}_{\mathrm{sH}}$ and $g^{\mathrm{z}}_{\mathrm{sH}}$. In an NM, the polarization of the spin Hall current is transverse to both the charge current and spin flow directions, so $g^{\mathrm{x}}_{\mathrm{sH}}$ and $g^{\mathrm{y}}_{\mathrm{sH}}$ vanish. Therefore, we mostly focus on $g^{\mathrm{z}}_{\mathrm{sH}}$. $g^{\mathrm{x}}_{\mathrm{sH}}$ and $g^{\mathrm{y}}_{\mathrm{sH}}$ provide additional information in magnetic systems.

We only include out-of-equilibrium spin currents in the SHE. Our definition for the spin Hall conductance $g^z_{\mathrm{sH}}$ is
\begin{equation}\label{eq:spin_hall_cond_definition}
g^{\mathrm{z}}_{\mathrm{sH}} = \frac{4\pi}{(-e)} \frac{ (-I^{\mathrm{s},\mathrm{z}}_{2} + I^{\mathrm{s},\mathrm{z}}_{3}) }{2\Delta V} \, .
\end{equation}
Similar expressions can be found for $g^{\mathrm{x}}_{\mathrm{sH}}$ and $g^{\mathrm{y}}_{\mathrm{sH}}$. In Eq.\, \eqref{eq:spin_hall_cond_definition}, $-I^{\mathrm{s},\mathrm{z}}_{2}$ is the spin current flowing into the top lead, and $I^{\mathrm{s},\mathrm{z}}_{3}$ flows from the bottom lead into the scattering region. The distribution of the electric potentials and the spin Hall current definition in Eq. \eqref{eq:spin_hall_cond_definition} result in {\it pure} spin currents in the top ($l=2$) and bottom ($l=3$) leads. More asymmetric geometries and heavy disorder can induce finite charge currents in the top and bottom leads, which we do not consider. By using Eqs. \eqref{eqs:spinCondsOutIn} and \eqref{eq:spin_current_LB_definition}, the spin Hall conductance defined in Eq. \eqref{eq:spin_hall_cond_definition} takes the explicit form
\begin{equation}
g_{\mathrm{sH}} = \frac{1}{4}\big[-(g^{\mathrm{in}}_{2,0} - g^{\mathrm{in}}_{2,1}) + (g^{\mathrm{in}}_{3,0} - g^{\mathrm{in}}_{3,1})\big] \, ,
\end{equation}
where we omit the label for the spin polarization direction.

We consider a tight-binding model on a square lattice with lattice constant $a$. The scattering region is quadratic and of area $L^{2}$ in terms of the length $L=Na$, where $N$ is the number of sites in one direction. The four leads are attached to each side of the scattering region. The width of each lead is $L$. The RSOC and localized AF spins are present only in the scattering region. The leads are NMs described by the nearest-neighbor hopping parameter $t$.

We denote operators and unit vectors with a hat. We define the operator
\begin{equation}
\hat{c}^{\dagger}_{\boldsymbol{r}} = \begin{pmatrix} \hat{c}^{\dagger}_{\boldsymbol{r}\uparrow} & \hat{c}^{\dagger}_{\boldsymbol{r}\downarrow}\end{pmatrix}  \,
\end{equation}
in terms of the creation (annihilation) operator $\hat{c}^{\dagger}_{\boldsymbol{r}s}$ ($\hat{c}_{\boldsymbol{r}s}$) for an electron at position $\boldsymbol{r}$ with spin $s=\uparrow$ or $s=\downarrow$. The itinerant spin density operator is $\boldsymbol{\hat{s}}_{\boldsymbol{r}} =(\hbar/2)\hat{c}^{\dagger}_{\boldsymbol{r}}\boldsymbol{\sigma}\hat{c}_{\boldsymbol{r}}$, where $\boldsymbol{\sigma} = \left( \sigma_{\mathrm{x}}, \sigma_{\mathrm{y}}, \sigma_{\mathrm{z}}\right)$ is the vector of Pauli matrices.

In the scattering region, the itinerant electrons are described by the real-space Hamiltonian $\hat{H} = \sum_{\boldsymbol{r}}\hat{H}_{\boldsymbol{r}}$, with
\begin{align}
\hat{H}_{\boldsymbol{r}} =& -t \sum_{\boldsymbol{\delta} =\pm\boldsymbol{\delta}_{\mathrm{x}},\pm\boldsymbol{\delta}_{\mathrm{y}} } \hat{c}^{\dagger}_{\boldsymbol{r}}\hat{c}_{\boldsymbol{r}+\boldsymbol{\delta}}     -J_{\mathrm{sd}} \boldsymbol{\mathcal{S}}_{\boldsymbol{r}}\cdot \boldsymbol{\hat{s}}_{\boldsymbol{r}} + \mathcal{V}^{\mathrm{imp}}_{\boldsymbol{r}}\hat{c}^{\dagger}_{\boldsymbol{r}}\hat{c}_{\boldsymbol{r}}  \nonumber \\
&+ \frac{\lambda_{\mathrm{R}}}{2a}  \big[ i(\hat{c}^{\dagger}_{\boldsymbol{r}}\sigma_{\mathrm{y}}\hat{c}_{\boldsymbol{r} + \boldsymbol{\delta}_{\mathrm{x}} } - \hat{c}^{\dagger}_{\boldsymbol{r}}\sigma_{\mathrm{x}}\hat{c}_{\boldsymbol{r} + \boldsymbol{\delta}_{\mathrm{y}} }) + \mathrm{H.c.}    \big] \, ,
\label{eq:hamiltonian_real_space}
\end{align}
where $t$ is the hopping energy and the sum ($\boldsymbol{\delta} =\pm\boldsymbol{\delta}_{\mathrm{x}},\pm\boldsymbol{\delta}_{\mathrm{y}}$) is over nearest neighbors, with $\boldsymbol{\delta}$ indicating hopping across distances $\boldsymbol{\delta}_{\mathrm{x}}=a\hat{x}$ and $\boldsymbol{\delta}_{\mathrm{y}}=a\hat{y}$ in the two spatial directions. In Eq.\, \eqref{eq:hamiltonian_real_space}, $J_{\mathrm{sd}}$ parametrizes the exchange coupling between the localized spins $\boldsymbol{\mathcal{S}}_{\boldsymbol{r}} $ and the itinerant spins $\boldsymbol{\hat{s}}_{\boldsymbol{r}}$. The localized spins are classical and static spins. $\mathcal{V}^{\mathrm{imp}}_{\boldsymbol{r}}$ is an elastic potential that models disorder. The statistical properties of $\mathcal{V}^{\mathrm{imp}}_{\boldsymbol{r}}$ are specified in Sec. \ref{section:numerics}. The second line in Eq. \eqref{eq:hamiltonian_real_space} is the RSOC due to the broken symmetry in the $\hat{z}$-direction, with $\lambda_{\mathrm{R}}$ being the strength of the RSOC and H.c. the Hermitian conjugate.

We consider AFs where the localized spins $\boldsymbol{\mathcal{S}}_{\boldsymbol{r}}$ are collinear and of equal length such that $|\boldsymbol{\mathcal{S}}_{\boldsymbol{r}}| \equiv \mathcal{S}$. However, the directions of the spins $\boldsymbol{\mathcal{S}}_{\boldsymbol{r}}$ are opposite for nearest neighbors, similar to a checkerboard pattern (as illustrated in Fig.\ \ref{Fig:fourterminalAF}). The explicit form is $\boldsymbol{\mathcal{S}}_{\boldsymbol{r}}= \mathcal{S}(-1)^{\frac{x}{a}}(-1)^{\frac{y}{a}}\boldsymbol{n}$, where the unit vector $\boldsymbol{n}$ is the Neel order parameter.

In the following, we use the dimensionless quantity
\begin{equation}
\xi_{\mathrm{sd}} = \frac{J_{\mathrm{sd}}\mathcal{S}\hbar}{2t} \,
\end{equation}
to parametrize the strength of the exchange coupling and the dimensionless variable
\begin{equation}
\xi_{\mathrm{R}} = \frac{\lambda_{\mathrm{R}}}{2at} \,
\end{equation}
to parametrize the strength of the RSOC.

We solve the scattering problem and obtain all conductances by utilizing the Python package KWANT \cite{groth2014kwant}.

\subsection{Normal Metals}

For comparison, we first consider an NM with RSOC, as studied in similar models \cite{pareekPRB2002may,hankipiPRBdec2004,ShengPRL2005buttiker,nikolicPRBaug2005}, which is well captured by our model (i.e., when $\xi_{\mathrm{sd}} = 0$). The Berry curvature contribution to the NM spin Hall conductivity is a universal value in certain regimes \cite{sinova2015spin}. However, the NM spin Hall conductance in mesoscopic systems connected to leads usually depends on the Fermi energy $E_{\mathrm{F}}$, system geometry, boundary conditions and RSOC.

On our square lattice, the Fermi energy lies between $-4t$ and $4t$. With our definition of the spin Hall conductance (Eq. \eqref{eq:spin_hall_cond_definition}), combined with the boundary conditions for the voltages, $g^{\mathrm{z}}_{\mathrm{sH}}$ is finite for the NM with RSOC, while the two other spin polarization components vanish. The NM spin Hall conductance as a function of Fermi energy is antisymmetric about $E_{\mathrm{F}} = 0$. In ballistic systems, the spin Hall conductance $g^{\mathrm{z}}_{\mathrm{sH}}$ increases with increasing system size (i.e., scales with the number of lead modes $N$).

We consider NMs with disorder in Sec. \ref{sec:disorderdNMs}. The SHE survives in NMs with weak spin-conserving disorder.

\subsection{Ballistic Antiferromagnets}\label{sec_sub_AFs_ballistic}

The spin Hall conductance $g_{\mathrm{sH}}$ of an AF with RSOC shows a richer behavior. The new features are caused by the localized spins described by the direction of the Neel order, $\boldsymbol{n}$. We mainly focus on AFs where the Neel order is out-of-plane ($\boldsymbol{n}=\pm\hat{z}$). First, in this section (Sec. \ref{sec_sub_AFs_ballistic}), we consider systems without disorder.

The coupling strength between localized spins and itinerant spins is governed by $\xi_{\mathrm{sd}}$. The other parameters that affect the spin Hall conductance are the Fermi energy $E_{\mathrm{F}}$, the RSOC strength $\xi_{\mathrm{R}}$, and the system size, determined by $N$, the number of transverse sites.

Odd and even effects (small oscillations) as a function of the system size $N$ are found. When $N$ is odd, the total number of localized spins is also odd such that the total magnetization is finite. In contrast, when $N$ is even, the total magnetization vanishes. Focusing on the main features, we only consider systems with even $N$, i.e., AFs with zero total magnetization, in the following.

We consider AFs with localized out-of-plane Neel order, $\boldsymbol{n}=\hat{z}$. In such systems, the $x$- and $y$-components of the spin Hall conductance vanish (similar to in an NM); we only discuss $g^{\mathrm{z}}_{\mathrm{sH}}$ here.
\begin{figure}[h]
	\centering
	\includegraphics[width=1.0\linewidth]{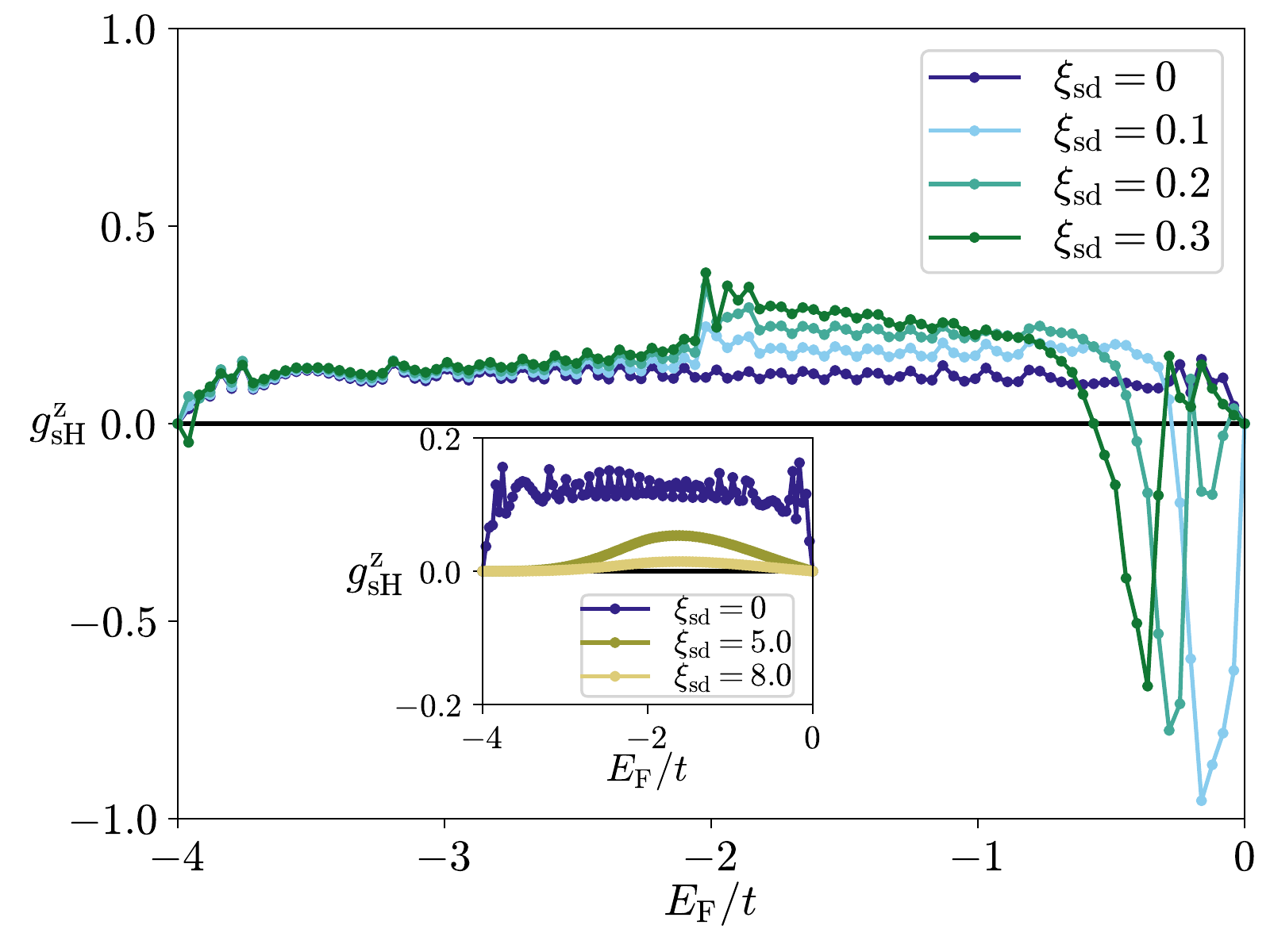}
\caption{Spin Hall conductance $g^{\mathrm{z}}_{\mathrm{sH}}$ as a function of Fermi energy $E_{\mathrm{F}}$ for an AF with out-of-plane Neel order, $\boldsymbol{n}=\hat{z}$. The system size is $N^{2}=100^{2}$, and the RSOC is $\xi_{\mathrm{R}}=0.1$. The results are for systems with increasing exchange coupling $\xi_{\mathrm{sd}}$. $\xi_{\mathrm{sd}}=0$ corresponds to an NM. The inset shows similar results for larger exchange parameters. }
	\label{Fig:AFexes:and:inset}
\end{figure}

In Fig. \ref{Fig:AFexes:and:inset}, we show how the exchange coupling $\xi_{\mathrm{sd}}$ influences the spin Hall conductance $g^{\mathrm{z}}_{\mathrm{sH}}$ as a function of the Fermi energy. The other system parameters are $\xi_{\mathrm{R}}=0.1$ and $N=100$. The spin Hall conductance is antisymmetric about $E_{\mathrm{F}}=0$. The general trend is that intermediate exchange couplings can dramatically increase the typical value and even change the sign of the spin Hall conductance. The increase in the typical value of the spin Hall conductance persists as long as the exchange coupling is not too strong, when $\xi_{\mathrm{sd}}$ is on the order of 1 or less (similar to the hopping energy $t$). The inset in Fig. \ref{Fig:AFexes:and:inset} demonstrates that when the exchange coupling is strong, the spin conductance $g^{\mathrm{z}}_{\mathrm{sH}}$ approaches zero because the system starts to become a poor conductor and the RSOC is weak with respect to the periodic potential. Variations in the exchange coupling $\xi_{\mathrm{sd}}$ influence the spin Hall conductance $g^{\mathrm{z}}_{\mathrm{sH}}$ much more than variations in the RSOC $\xi_{\mathrm{R}}$ (not shown).
\begin{figure}[h]
	\centering
	\includegraphics[width=1.0\linewidth]{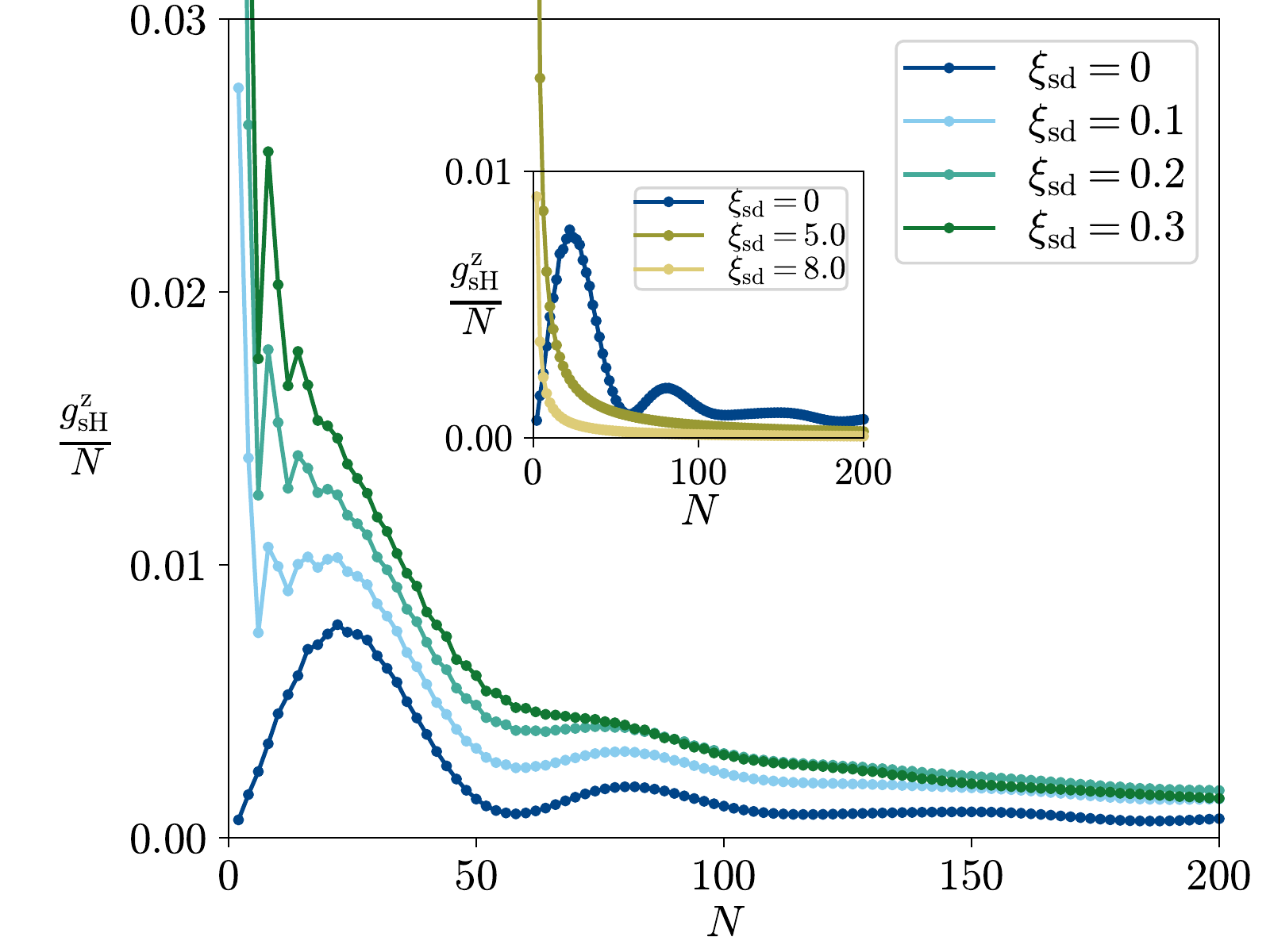}
\caption{Spin Hall conductance $g^{\mathrm{z}}_{\mathrm{sH}}/N$ as a function of length $N$, where $N$ is even. The Neel orientation is $\boldsymbol{n}=\hat{z}$. The RSOC is $\xi_{\mathrm{R}}=0.1$, and the Fermi energy is $E_{\mathrm{F}}=-2t$. The inset shows similar results for larger exchange couplings. }
	\label{Fig:AF:shVsSizeDivN}
\end{figure}

In the square ballistic system, the spin Hall conductance increases as the width of the leads $L=Na$ increases because the number of modes increases. A measure of the spin Hall conductance per mode is $g^{\mathrm{z}}_{\mathrm{sH}}/N$. For small systems (small $N$), finite size effects occur. With increasing $N$, one would expect $g^{\mathrm{z}}_{\mathrm{sH}}/N$ to approach a constant value. In Fig. \ref{Fig:AF:shVsSizeDivN}, we show $g^{\mathrm{z}}_{\mathrm{sH}}/N$ as a function of $N$ (with $N$ even). The curves in Fig. \ref{Fig:AF:shVsSizeDivN} correspond to several exchange couplings $\xi_{\mathrm{sd}}$, while the remaining parameters are $\boldsymbol{n}=\hat{z}$, $\xi_{\mathrm{R}}=0.1$ and $E_{\mathrm{F}} = -2t$. For the relevant parameters, Fig. \ref{Fig:AF:shVsSizeDivN} shows that $g^{\mathrm{z}}_{\mathrm{sH}}/N$ approaches a constant when $N$ is on the order of 100 and greater. To minimize finite size effects while keeping the computation time moderate, we mostly focus on systems with $N=100$ or larger.

We also consider a few cases (results not shown) where the localized spins are rotated in-plane ($\boldsymbol{n}\neq \pm \hat{z}$). Rotating the Neel order $\boldsymbol{n}$ in-plane modifies the amplitude of $g^{\mathrm{z}}_{\mathrm{sH}}$. Additionally, in-plane localized spins can induce finite components of $g^{\mathrm{x}}_{\mathrm{sH}}$ and $g^{\mathrm{y}}_{\mathrm{sH}}$, related to the spin polarizations in the $x$- and $y$-directions, respectively.


\section{Intrinsic Spin Hall Conductivity}\label{section:kubo}

We complement our numerical results of the spin Hall conductance related to a finite scattering region with analytical results of the spin Hall conductivity in infinite systems. An electric field $\boldsymbol{\mathcal{E}} = \mathcal{E}_{\mathrm{x}}\hat{x}$ in the $x$-direction induces spin current densities $j^{\mathrm{s},\mathrm{x}}_{\mathrm{y}}$, $j^{\mathrm{s},\mathrm{y}}_{\mathrm{y}}$, and $j^{\mathrm{s},\mathrm{z}}_{\mathrm{y}}$, which flow in the $y$-direction with spin polarization along the $x$-, $y$-, and $z$-directions, respectively. Correspondingly, the spin Hall conductivity has three polarization components: $\sigma^{\mathrm{x}}_{\mathrm{sH}}$, $\sigma^{\mathrm{y}}_{\mathrm{sH}}$, and $\sigma^{\mathrm{z}}_{\mathrm{sH}}$. The dimensionless spin Hall conductivity $\sigma^{\mathrm{z}}_{\mathrm{sH}}$ related to the spin polarization along $z$ follows from
\begin{equation}\label{eq:definitionSpinHallConductivitySigmaAndE}
j^{\mathrm{s},\mathrm{z}}_{\mathrm{y}} = \frac{-e}{4\pi}  \sigma^{\mathrm{z}}_{\mathrm{sH}}\mathcal{E}_{\mathrm{x}} \, ,
\end{equation}
and similarly for the two other spin polarization components. In Eq. \eqref{eq:definitionSpinHallConductivitySigmaAndE}, $j^{\mathrm{s},\mathrm{z}}_{\mathrm{y}} = j^{\mathrm{s},\mathrm{z}}_{\mathrm{y}}(\omega\rightarrow 0)$ is the average spin current, which we evaluate in the static limit when the frequency $\omega\rightarrow 0$. Eq. \eqref{eq:TimeDomainSpinCurrentDefs} defines the average spin current in the time domain, which yields $j^{\mathrm{s},\mathrm{z}}_{\mathrm{y}}(\omega)$ after a Fourier transform to the frequency domain.

In disordered systems, there are many contributions to the spin Hall conductivity. One of these contributions, the Berry phase term within the linear response Kubo formula, is intrinsic and independent of the mean free path. In this section, we calculate the intrinsic spin Hall conductivity $\sigma^{\mathrm{z}}_{\mathrm{sH}}$ for an AF with RSOC.

\subsection{Diagonalization of the Hamiltonian of a Bulk Antiferromagnet}

We consider an AF with RSOC subject to periodic boundary conditions in two spatial directions. The intrinsic contribution to the spin Hall conductivity is calculated in the absence of scattering. The itinerant electrons in the AF are described by the Hamiltonian \eqref{eq:hamiltonian_real_space} with no disorder scattering, $\mathcal{V}^{\mathrm{imp}}_{\boldsymbol{r}} = 0$. The system has $N^{2}$ (with $N$ even) localized spins. Nearest-neighbor localized spins point in opposite directions, as shown in Fig.\ \ref{Fig:fourterminalAF}.

\begin{figure}[h]
	\centering
	\includegraphics[width=0.7\linewidth]{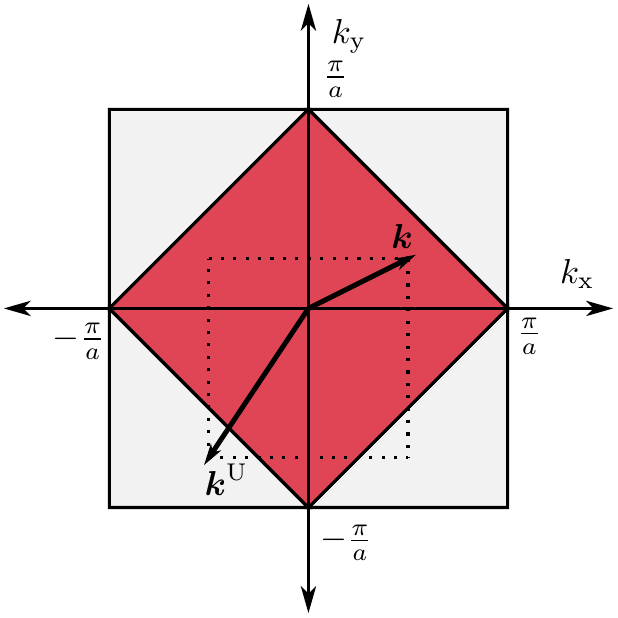}
\caption{Momentum space. The magnetic Brillouin zone (MBZ) for the AF is the red region. The vectors illustrate a momentum $\boldsymbol{k}$ and its umklapp momentum $\boldsymbol{k}^{\mathrm{U}}$. When $\boldsymbol{k}$ lies in the MBZ, $\boldsymbol{k}^{\mathrm{U}}$ is outside the MBZ, and vice versa. The umklapp momentum is $\boldsymbol{k}^{\mathrm{U}} = \boldsymbol{k} -(\frac{\pi}{a}, \frac{\pi}{a})$, as indicated by the dotted lines. Similar considerations apply to $\boldsymbol{k}^{\mathrm{U}}$ when $\boldsymbol{k}$ lies in one of the other triangles/quadrants. }
	\label{Fig:mbzAndUmklapp}
\end{figure}

To diagonalize the AF Hamiltonian \eqref{eq:hamiltonian_real_space}, we Fourier transform the annihilation operators
\begin{equation}
\hat{c}_{\boldsymbol{r}}=\frac{1}{N} \sum_{\boldsymbol{k}}\mathrm{Exp}(i\boldsymbol{k}\cdot\boldsymbol{r})\hat{c}_{\boldsymbol{k}} \, .
\end{equation}
In Fourier space, the tight-binding hopping part of the energy dispersion is
\begin{equation}\label{eq:TBhoppingEnergyMomentumSpace}
\varepsilon_{0}(\boldsymbol{k}) = -2t (\cos k_{\mathrm{x}}a + \cos k_{\mathrm{y}}a) \, .
\end{equation}
In the AF, the sd exchange coupling results in pairwise coupling between different momenta: the momentum $\boldsymbol{k}$ and its umklapp momentum $\boldsymbol{k}^{\mathrm{U}}$, as illustrated in Fig.\ \ref{Fig:mbzAndUmklapp}.

To capture the umklapp scattering in a $4 \times 4$ Hamiltonian, we introduce the operator $\hat{\mathcal{C}}^{\dagger}_{\boldsymbol{k}} = \begin{pmatrix} \hat{c}^{\dagger}_{\boldsymbol{k}\uparrow} & \hat{c}^{\dagger}_{\boldsymbol{k}\downarrow} & \hat{c}^{\dagger}_{\boldsymbol{k}^{\mathrm{U}}\uparrow} & \hat{c}^{\dagger}_{\boldsymbol{k}^{\mathrm{U}}\downarrow}   \end{pmatrix} $, and similarly for $\hat{\mathcal{C}}_{\boldsymbol{k}}$. In this $4 \times 4$ basis, the Fourier transform of the AF Hamiltonian \eqref{eq:hamiltonian_real_space} becomes
\begin{align}\label{eq:AFhamiltonianMomentumSpace}
\hat{\mathcal{H}} = \sum_{\boldsymbol{k}\in\Diamond} \hat{\mathcal{C}}^{\dagger}_{\boldsymbol{k}} \Bigg[ \varepsilon_{0}(\boldsymbol{k}) + \begin{pmatrix}  \mathbb{H}_{\mathrm{R}}  & \mathbb{H}_{\mathrm{sd}} \\  \mathbb{H}_{\mathrm{sd}} &   \mathbb{H}_{\mathrm{R}} \end{pmatrix} \Bigg] \hat{\mathcal{C}}_{\boldsymbol{k}}  \,
\end{align}
in terms of the $2 \times 2$ matrices $\mathbb{H}_{\mathrm{R}} = 2t\xi_{\mathrm{R}} (\sigma_{\mathrm{x}}\sin k_{\mathrm{y}}a -\sigma_{\mathrm{y}}\sin k_{\mathrm{x}}a )$ and $\mathbb{H}_{\mathrm{sd}} = -t\xi_{\mathrm{sd}}(\boldsymbol{n}\cdot\boldsymbol{\sigma})$. In Eq.\ \eqref{eq:AFhamiltonianMomentumSpace}, the sum is over $\boldsymbol{k}\in\Diamond$, the momenta $\boldsymbol{k}$ within the {\it magnetic} Brillouin zone, as shown in Fig.\ \ref{Fig:mbzAndUmklapp}.

The Hamiltonian in Eq.\ \eqref{eq:AFhamiltonianMomentumSpace} is diagonalized by a unitary matrix $\mathcal{U}_{\boldsymbol{k}}$ that transforms the operators as $\hat{\eta}_{\boldsymbol{k}} =  \mathcal{U}^{\dagger}_{\boldsymbol{k}} \mathcal{\hat{C}}_{\boldsymbol{k}} $ and $\hat{\eta}^{\dagger}_{\boldsymbol{k}} = \mathcal{\hat{C}}^{\dagger}_{\boldsymbol{k}} \mathcal{U}_{\boldsymbol{k}} $. The resulting diagonal Hamiltonian is
\begin{equation}\label{eq:diagonalAFhamiltonianMomentumSpace}
\hat{\mathcal{H}} = \sum_{\boldsymbol{k}\in \Diamond}\hat{\eta}^{\dagger}_{\boldsymbol{k}} \big[\varepsilon_{0}(\boldsymbol{k}) + \mathcal{D}_{\boldsymbol{k}} \big] \hat{\eta}_{\boldsymbol{k}} \, ,
\end{equation}
where $\mathcal{D}_{\boldsymbol{k}}$ is a diagonal matrix containing four eigenvalues. We specify the eigenvectors and eigenenergies for three cases: out-of-plane localized spins, AF spins rotating in the $x$-$z$-plane, and AF spins rotating in the $y$-$z$-plane, as summarized in App. \ref{appendix:antiferromagnet}.

\subsection{Kubo Formula}\label{sec:sub:Kubo}

We calculate the intrinsic part of the spin Hall conductivity by considering the average spin current in linear response to an applied electric field. A time-dependent electric field $\mathcal{E}_{\mathrm{x}}(t) = -\partial_{t}\mathcal{A}_{\mathrm{x}}(t)$ exists in the $x$-direction, expressed in terms of the vector potential $\mathcal{A}_{\mathrm{x}}$, which oscillates with frequency $\omega$ but does not vary with position $\boldsymbol{r}$. In this gauge, the scalar electric potential is zero. We evaluate the spin Hall conductivity in the static limit, $\omega\rightarrow 0 $.

Consider the Hamiltonian $\hat{H}$ in Eq. \eqref{eq:hamiltonian_real_space} with $\mathcal{V}^{\mathrm{imp}}_{\boldsymbol{r}}=0$. The minimal coupling to the vector potential $\mathcal{A}_{\mathrm{x}}$ changes the corresponding momentum as $p_{\mathrm{x}} \rightarrow p_{\mathrm{x}} + e\mathcal{A}_{\mathrm{x}}$, which, in the tight-binding model, changes the hoppings in the $x$-direction. We use the Peierls substitution \cite{PRBgrafVoglPeierls} to express the change in the hoppings, which, for example, yields $\hat{c}^{\dagger}_{\boldsymbol{r}+\boldsymbol{\delta}_{\mathrm{x}}}\hat{c}_{\boldsymbol{r}} \rightarrow \hat{c}^{\dagger}_{\boldsymbol{r}+\boldsymbol{\delta}_{\mathrm{x}}}\hat{c}_{\boldsymbol{r}}\mathrm{Exp} \big[i(-e/\hbar)\int_{\boldsymbol{r}}^{\boldsymbol{r}+\boldsymbol{\delta}_{\mathrm{x}}} \boldsymbol{\mathcal{A}}(t)\cdot d\boldsymbol{r}'   \big] = \hat{c}^{\dagger}_{\boldsymbol{r}+\boldsymbol{\delta}_{\mathrm{x}}}\hat{c}_{\boldsymbol{r}}\mathrm{Exp} \big[i(-e/\hbar)a\mathcal{A}_{\mathrm{x}}(t)  \big] $. As a consequence, the unperturbed Hamiltonian $\hat{H}$ transforms into $\hat{H}\rightarrow \hat{H} + \hat{\delta H}(t)$. To first order in the vector potential, the variation in the Hamiltonian is
\begin{equation}\label{eq:shiftHamiltonianCurrentOp}
\hat{\delta H}(t) = - a \sum_{\boldsymbol{r}} \hat{j}^{\mathrm{c}}_{\boldsymbol{r},\boldsymbol{\delta}_{\mathrm{x}}} \mathcal{A}_{\mathrm{x}}(t)  \,
\end{equation}
in terms of the charge current operator $\hat{j}^{\mathrm{c}}_{\boldsymbol{r},\boldsymbol{\delta}_{\mathrm{x}}}$ between sites $\boldsymbol{r}$ and $\boldsymbol{r}+\boldsymbol{\delta}_{\mathrm{x}}$. The charge current operator is separated into
\begin{equation}\label{eq:theDefinitionChargeCurrentOperator}
\hat{j}^{\mathrm{c}}_{\boldsymbol{r},\boldsymbol{\delta}_{\mathrm{x}}}  =   \hat{j}^{\mathrm{c,}0}_{\boldsymbol{r},\boldsymbol{\delta}_{\mathrm{x}}}  + \hat{j}^{\mathrm{c,a}}_{\boldsymbol{r},\boldsymbol{\delta}_{\mathrm{x}}}  \, ,
\end{equation}
where the first term is the normal charge current contribution and the second, anomalous, term is caused by the spin-orbit interaction
\begin{subequations}\label{eq:defChargeCurrOpTwoTerms}
\begin{align}
\hat{j}^{\mathrm{c,}0}_{\boldsymbol{r},\boldsymbol{\delta}_{\mathrm{x}}} =& -e\frac{it}{\hbar}(\hat{c}^{\dagger}_{\boldsymbol{r}+\boldsymbol{\delta}_{\mathrm{x}}}\hat{c}_{\boldsymbol{r}} - \hat{c}^{\dagger}_{\boldsymbol{r}} \hat{c}_{\boldsymbol{r}+\boldsymbol{\delta}_{\mathrm{x}}} )  \, , \\
\hat{j}^{\mathrm{c,a}}_{\boldsymbol{r},\boldsymbol{\delta}_{\mathrm{x}}} =& e\xi_{\mathrm{R}}\frac{t}{\hbar} (\hat{c}^{\dagger}_{\boldsymbol{r}+\boldsymbol{\delta}_{\mathrm{x}}}\sigma_{\mathrm{y}}\hat{c}_{\boldsymbol{r}} + \hat{c}^{\dagger}_{\boldsymbol{r}}\sigma_{\mathrm{y}}\hat{c}_{\boldsymbol{r}+\boldsymbol{\delta}_{\mathrm{x}}})  \, .
\end{align}
\end{subequations}
The electric charge is conserved such that Eqs. \eqref{eq:shiftHamiltonianCurrentOp}, \eqref{eq:theDefinitionChargeCurrentOperator} and \eqref{eq:defChargeCurrOpTwoTerms} (together with the charge continuity equations Eqs. \eqref{eq:continuityEqChargeGeneral}, \eqref{eq:chargeCurrentsTwoTerms} and \eqref{eq:chargeCurrentsExplicit}) yield a well-defined charge current.

The spin is not conserved in a system with RSOC. In the appendix, we derive a spin continuity equation  by considering the temporal rate of change of the spin density operator $\hat{\boldsymbol{s}}_{\boldsymbol{r}}$. The spin continuity equation (Eq. \eqref{eq:spinContinuityEquationFull}) includes nonconserving terms due to the RSOC and spin-transfer torques, as summarized in Eqs. \eqref{eq:spinContinuityEquationFull}, \eqref{eq:appTwoTypesSpinCurrent}, \eqref{eq:spinCurrentNormalHoppings}, \eqref{eq:SpinCurrentsSocX}, \eqref{eq:SpinCurrentsSocY} and \eqref{eq:SpinCurrentsSocZ} in App. \ref{appendix:TBcurrents}. In the bulk of a system with RSOC, the definition of the spin current is ambiguous. We use a conventional definition of the spin current as in Ref.\ \cite{sinovaUniversalPRL2004} for comparison of our results with the case of an NM (semiconductor).

Our definition of the spin current operator, which describes flow in the $y$-direction with spin polarization along $z$, is
\begin{equation}\label{eq:spinCurrentFlowYPolZdef}
\hat{j}^{\mathrm{s,0,z}}_{\boldsymbol{r},\boldsymbol{\delta}_{\mathrm{y}}} =  \frac{it}{2}(\hat{c}^{\dagger}_{\boldsymbol{r}+\boldsymbol{\delta}_{\mathrm{y}}}\sigma_{\mathrm{z}}\hat{c}_{\boldsymbol{r}} - \hat{c}^{\dagger}_{\boldsymbol{r}}\sigma_{\mathrm{z}}\hat{c}_{\boldsymbol{r}+\boldsymbol{\delta}_{\mathrm{y}}})  \, ,
\end{equation}
where $\hat{j}^{\mathrm{s,0,z}}_{\boldsymbol{r},\boldsymbol{\delta}_{\mathrm{y}}}$ is the spin current between sites $\boldsymbol{r}$ and $\boldsymbol{r}+\boldsymbol{\delta}_{\mathrm{y}}$. The spin current in Eq. \eqref{eq:spinCurrentFlowYPolZdef} stems from Eqs. \eqref{eq:spinContinuityEquationFull}, \eqref{eq:appTwoTypesSpinCurrent} and \eqref{eq:spinCurrentNormalHoppings}.

In the Kubo formula for the spin Hall conductivity $\sigma^{\mathrm{z}}_{\mathrm{sH}}$, we consider the average spin current operator
\begin{equation}\label{eq:defAvSpinCurrentDensity}
\hat{j}^{\mathrm{s,z}}_{\mathrm{y}}  =\frac{1}{(Na)^{2}}a\sum_{\boldsymbol{r}}\hat{j}^{\mathrm{s,0,z}}_{\boldsymbol{r},\boldsymbol{\delta}_{\mathrm{y}}}  \, ,
\end{equation}
where $N^{2}$ is the total number of lattice sites. The spin current $j^{\mathrm{s,z}}_{\mathrm{y}}$ that defines $\sigma^{\mathrm{z}}_{\mathrm{sH}}$ [Eq. \eqref{eq:definitionSpinHallConductivitySigmaAndE}] now follows from the expectation value of the correlation function between the average spin current operator $\hat{j}^{\mathrm{s,z}}_{\mathrm{y}}$ and the charge current operator defined in Eq. \eqref{eq:shiftHamiltonianCurrentOp}. Within the linear response, the average spin current $j^{\mathrm{s,z}}_{\mathrm{y}}(t)$ in the time domain is
\begin{align}\label{eq:TimeDomainSpinCurrentDefs}
j^{\mathrm{s,z}}_{\mathrm{y}}(t) =&\frac{1}{(Na)^{2}} \int_{-\infty}^{\infty}dt' \Theta(t-t')\frac{(-i)}{\hbar}  \mathcal{A}_{\mathrm{x}}(t')  \nonumber \\
& \times\big\langle \big[ a\sum_{\boldsymbol{r}}\hat{j}^{\mathrm{s,0,z}}_{\boldsymbol{r},\boldsymbol{\delta}_{\mathrm{y}}}(t) , -a\sum_{\boldsymbol{r}'} \hat{j}^{\mathrm{c}}_{\boldsymbol{r}',\boldsymbol{\delta}_{\mathrm{x}}}(t')   \big] \big\rangle_{\mathrm{eq.}}  \, ,
\end{align}
where $\Theta(t-t')$ is the Heaviside-theta function, the operators have time dependence in the Heisenberg picture, and the expectation value of the commutator is evaluated for the equilibrium many-particle state.

We express the spin Hall conductivity in the $4\times4$ basis using the single-particle eigenstates $\psi_{\boldsymbol{k}n}$ along with their corresponding eigenenergies $E_{n}(\boldsymbol{k})$. The real part of the dimensionless spin Hall conductivity is
\begin{align}\label{eq:sHallConductivity4basisGeneral}
\sigma^{\mathrm{z}}_{\mathrm{sH}} =& \frac{4\pi}{(-e)} \frac{\hbar}{N^{2}}\sum_{\boldsymbol{k}\in\Diamond}\sum_{n,n'\neq n}(f_{n'}-f_{n})  \nonumber \\
& \times\frac{\mathrm{Im}\big\{  \psi^{\dagger}_{\boldsymbol{k}n'} \mathcal{J}^{\mathrm{s,0,z}}_{\mathrm{y}} \psi_{\boldsymbol{k}n}  \psi^{\dagger}_{\boldsymbol{k}n} \mathcal{J}^{\mathrm{c,a}}_{\mathrm{x}} \psi_{\boldsymbol{k}n'}  \big\} }{(E_{n}-E_{n'})^{2}}  \, ,
\end{align}
where $f_{n}=f_{\mathrm{FD}}(E_{n}(\boldsymbol{k})-\mu)$ in terms of the Fermi-Dirac distribution $f_{\mathrm{FD}}$, with $\mu$ being the chemical potential. In Eq. \eqref{eq:sHallConductivity4basisGeneral}, the $4\times4$ matrices $\mathcal{J}^{\mathrm{s,0,z}}_{\mathrm{y}}$ and $\mathcal{J}^{\mathrm{c,a}}_{\mathrm{x}}$ originate from the spin current $\hat{j}^{\mathrm{s,0,z}}_{\boldsymbol{r},\boldsymbol{\delta}_{\mathrm{y}}}$ and the charge current $\hat{j}^{\mathrm{c,}a}_{\boldsymbol{r},\boldsymbol{\delta}_{\mathrm{x}}}$, respectively (the contribution from $\hat{j}^{\mathrm{c,}0}_{\boldsymbol{r},\boldsymbol{\delta}_{\mathrm{x}}}$ vanishes), expressed as
\begin{subequations}
\begin{align}
\mathcal{J}^{\mathrm{s,0,z}}_{\mathrm{y}} =& (\sigma_{\mathrm{z}}\otimes \tau_{0})t\sin k_{\mathrm{y}}a     \, , \label{subeq:spinCurr4BasisMatrix} \\
\mathcal{J}^{\mathrm{c,a}}_{\mathrm{x}} =& (\sigma_{\mathrm{y}}\otimes \tau_{0})\frac{et}{\hbar}2\xi_{\mathrm{R}} \cos k_{\mathrm{x}}a \, ,
\end{align}
\end{subequations}
where $\tau_{0}$ is a $2\times2$ unit matrix.

We also use Eq. \eqref{eq:sHallConductivity4basisGeneral} to calculate $\sigma^{\mathrm{x}}_{\mathrm{sH}}$ and $\sigma^{\mathrm{y}}_{\mathrm{sH}}$ by replacing $\sigma_{\mathrm{z}}$ in the expression for the spin current $\mathcal{J}^{\mathrm{s,0,z}}_{\mathrm{y}}$ in Eq. \eqref{subeq:spinCurr4BasisMatrix} with $\sigma_{\mathrm{x}}$ or $\sigma_{\mathrm{y}}$, respectively.

\subsubsection{Out-of-plane localized spins}\label{seq:kuboOutofplaneSpins}

We now present results for the spin Hall conductivity of an AF with out-of-plane spins, $\boldsymbol{n} = \hat{z}$. This case involves two (doubly degenerate) AF eigenenergies: $E_{\pm}(\boldsymbol{k}) = \varepsilon_{0}(\boldsymbol{k}) + \Delta^{\mathrm{z}}_{\pm}(\boldsymbol{k})$, where the splitting of the two bands is
\begin{equation}\label{eq:twoAFeigenEnergiesOutOfPLane}
\Delta^{\mathrm{z}}_{\pm}(\boldsymbol{k}) = \pm t\sqrt{\xi_{\mathrm{sd}}^{2} + (2\xi_{\mathrm{R}})^{2} \left( \sin^{2} k_{\mathrm{x}}a + \sin^{2} k_{\mathrm{y}} a \right) } \, .
\end{equation}
The spin Hall conductivity for out-of-plane AF spins is
\begin{align}\label{eq:spinHconductOutPlaneSpinsExprs}
\sigma^{\mathrm{z}}_{\mathrm{sH}}  =& \frac{ (2\pi)^{2} }{ N^{2} /2 } \sum_{\boldsymbol{k}\in\Diamond}\frac{\cos k_{\mathrm{x}}a}{ 2\pi} \frac{ (f_{-}-f_{+}) }{ (\frac{\Delta^{\mathrm{z}}_{+}}{t})^{3} } (2\xi_{\mathrm{R}} )^{2}  \sin^{2} k_{\mathrm{y}}a  \, ,
\end{align}
where $f_{\pm}=f_{\mathrm{FD}}(E_{\pm}-\mu)$. We evaluate Eq. \eqref{eq:spinHconductOutPlaneSpinsExprs} in the zero-temperature limit, where $f_{\mathrm{FD}} (E_{\pm} - \mu) \rightarrow \Theta (E_{\mathrm{F}} - E_{\pm} )$. In the bulk limit, the number of lattice sites $N^{2}\rightarrow \infty$ such that $\sum_{\boldsymbol{k}\in\Diamond} \rightarrow (1/2) \big[(Na)/(2\pi)\big]^{2}\int_{\boldsymbol{k}\in\Diamond} d^{2}k$.
\begin{figure}[h]
	\centering
	\includegraphics[width=1.0\linewidth]{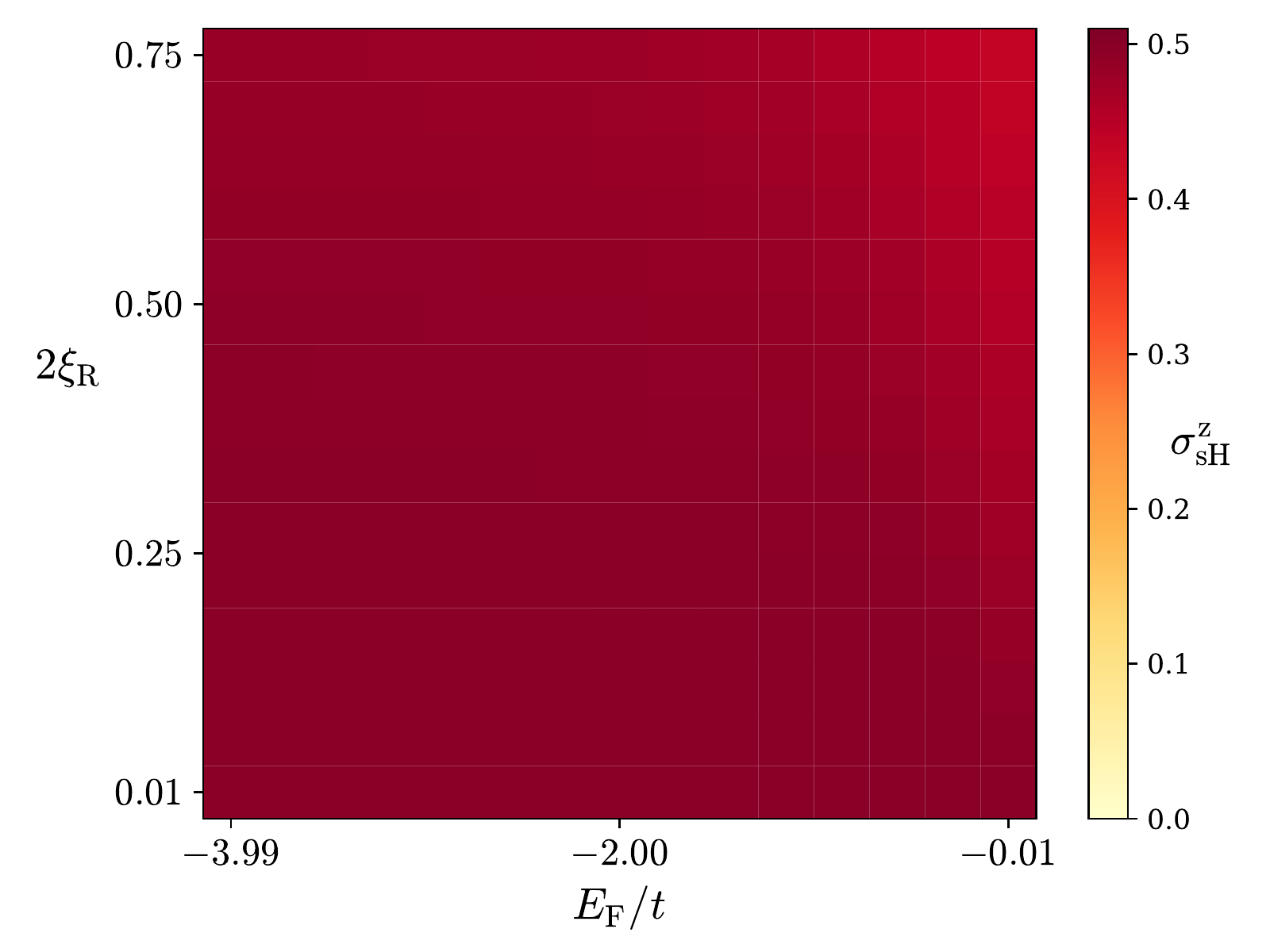}
\caption{Spin Hall conductivity $\sigma^{\mathrm{z}}_{\mathrm{sH}}$ in the NM limit ($\xi_{\mathrm{sd}} = 0$). Here, we numerically evaluate Eq. \eqref{eq:spinHconductOutPlaneSpinsExprs} for values of $E_{\mathrm{F}}$ and $2\xi_{\mathrm{R}}$ on an equidistant grid with $15\times15$ points. The spin Hall conductivity $\sigma^{\mathrm{z}}_{\mathrm{sH}}$ vanishes when $\xi_{\mathrm{R}}$ is exactly 0; however, in the limit $\sigma^{\mathrm{z}}_{\mathrm{sH}}(\xi_{\mathrm{R}}\rightarrow0)$, the amplitude changes in a discontinuous manner (not visible in the plot). }
	\label{Fig:kuboNormalMetall}
\end{figure}

The spin Hall conductivity $\sigma^{\mathrm{z}}_{\mathrm{sH}}$ (Eq. \eqref{eq:spinHconductOutPlaneSpinsExprs}) as a function of Fermi energy is antisymmetric about $E_{\mathrm{F}} = 0$. Furthermore, $\sigma^{\mathrm{z}}_{\mathrm{sH}}$ vanishes at $E_{\mathrm{F}} = \pm 4t$. For out-of-plane AF spins, the spin Hall conductivities related to the two other spin polarizations, $\sigma^{\mathrm{x}}_{\mathrm{sH}}$ and $\sigma^{\mathrm{y}}_{\mathrm{sH}}$, both vanish. The spin Hall conductivity is the same when $\boldsymbol{n} = -\hat{z}$ as when $\boldsymbol{n} = \hat{z}$. These considerations also apply to the spin Hall conductivity in the NM limit.

First, consider $\sigma^{\mathrm{z}}_{\mathrm{sH}}$ in the NM limit ($\xi_{\mathrm{sd}}=0$). As shown in Fig.\ \ref{Fig:kuboNormalMetall}, $\sigma^{\mathrm{z}}_{\mathrm{sH}}$ is finite and approximately $0.5$ when the Fermi energy $E_{\mathrm{F}}$ is between $-4t$ and $0$ and the RSOC is finite. The amplitude of $\sigma^{\mathrm{z}}_{\mathrm{sH}}$ does not vary greatly as a function of the RSOC. However, $\sigma^{\mathrm{z}}_{\mathrm{sH}}$ exactly vanishes when $\xi_{\mathrm{R}} = 0$. In Fig.\ \ref{Fig:kuboNormalMetall}, we plot $\sigma^{\mathrm{z}}_{\mathrm{sH}}$ when the Rashba parameter is positive, $\xi_{\mathrm{R}}>0$. The amplitude of $\sigma^{\mathrm{z}}_{\mathrm{sH}}$ is discontinuous in the limit $\xi_{\mathrm{R}}\rightarrow 0$. Our results for the tight-binding model agree with the fact that the dimensionless spin Hall conductivity is $\pm1/2$, the "universal" value \cite{sinovaUniversalPRL2004}, in the continuum model with linear-in-momentum RSOC (excluding vertex corrections).
\begin{figure}[h]
	\centering
	\includegraphics[width=1.0\linewidth]{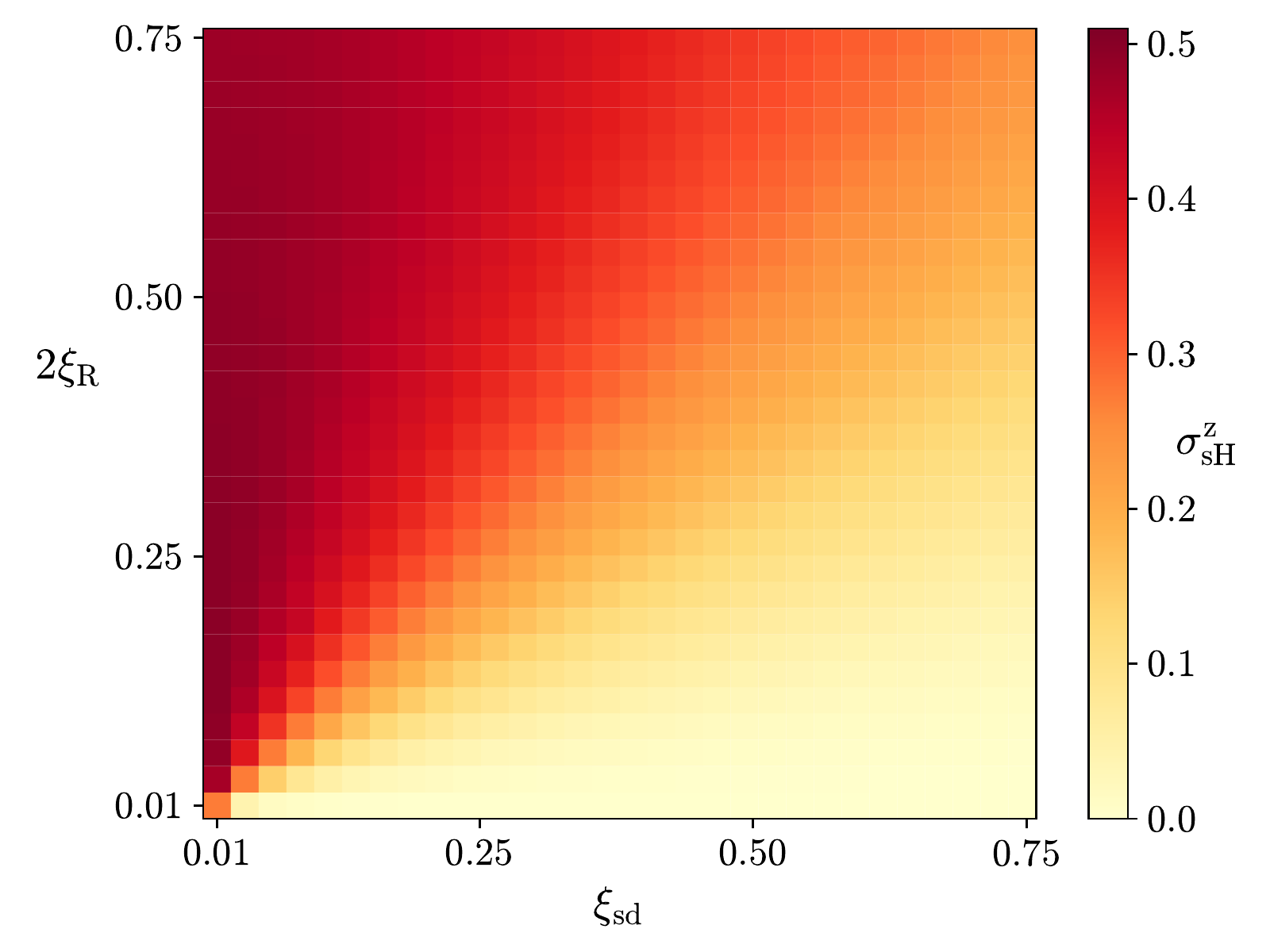}
\caption{Spin Hall conductivity $\sigma^{\mathrm{z}}_{\mathrm{sH}}$ as a function of exchange interaction $\xi_{\mathrm{sd}}$ and RSOC $2\xi_{\mathrm{R}}$ in an AF with out-of-plane spins, $\boldsymbol{n}=\hat{z}$. The Fermi energy is $E_{\mathrm{F}} = -2t$. Each square represents $\sigma^{\mathrm{z}}_{\mathrm{sH}}$ as calculated from Eq. \eqref{eq:spinHconductOutPlaneSpinsExprs}. }
	\label{Fig:kuboAF2Dplot}
\end{figure}

Next, we study the behavior of the spin Hall conductivity in AFs. We plot typical results for the AF spin Hall conductivity for various exchange couplings $\xi_{\mathrm{sd}}$ and RSOCs $2\xi_{\mathrm{R}}$ in Fig.\ \ref{Fig:kuboAF2Dplot}. Fig.\ \ref{Fig:kuboAF2Dplot} shows that the amplitude of $\sigma^{\mathrm{z}}_{\mathrm{sH}}$ is greatly reduced for an AF with a strong exchange interaction compared to a weak AF (or NM) with the same RSOC. Naturally, when the exchange interaction in the AF is weak, $\xi_{\mathrm{sd}}\approx0$, the spin Hall conductivity is close to 1/2, the value in the NM case. The reduction in the amplitude of $\sigma^{\mathrm{z}}_{\mathrm{sH}}$ as the exchange coupling $\xi_{\mathrm{sd}}$ increases is less dramatic when the RSOC is large. The results in Fig.\ \ref{Fig:kuboAF2Dplot} (where $E_{\mathrm{F}} = -2t$) are representative of the results of $\sigma^{\mathrm{z}}_{\mathrm{sH}}$ for Fermi energies $-4t < E_{\mathrm{F}} < 0$ because the numerical values and the shapes of the plots are very similar. The general trend is that the Berry phase contribution to the spin Hall conductivity decreases with increasing exchange interaction in AFs.

As will be shown in Sec. \ref{section:numerics}, an exact numerical calculation of the transport properties using the Landauer-B$\ddot{\text{u}}$ttiker formalism gives, in certain regimes, the opposite behavior; the spin Hall conductance increases with increasing exchange interaction. This is yet another example that interpretation of the spin Hall conductivity based on only the Berry phase contribution should be applied with great care.

\subsubsection{In-plane localized spins}\label{seq:kuboInplaneSpins}

In this section, we show how the spin Hall conductivity behaves when the AF spins rotate towards the in-plane configuration. We consider two cases. We use the polar angle $\theta$ to describe how the Neel order parameter $\boldsymbol{n}$ rotates in either the $x$-$z$-plane or the $y$-$z$-plane. We summarize the expressions for the spin Hall conductivities in the two scenarios in App. \ref{appendix:antiferromagnet}.

\begin{figure}[h]
	\centering
	\includegraphics[width=1.0\linewidth]{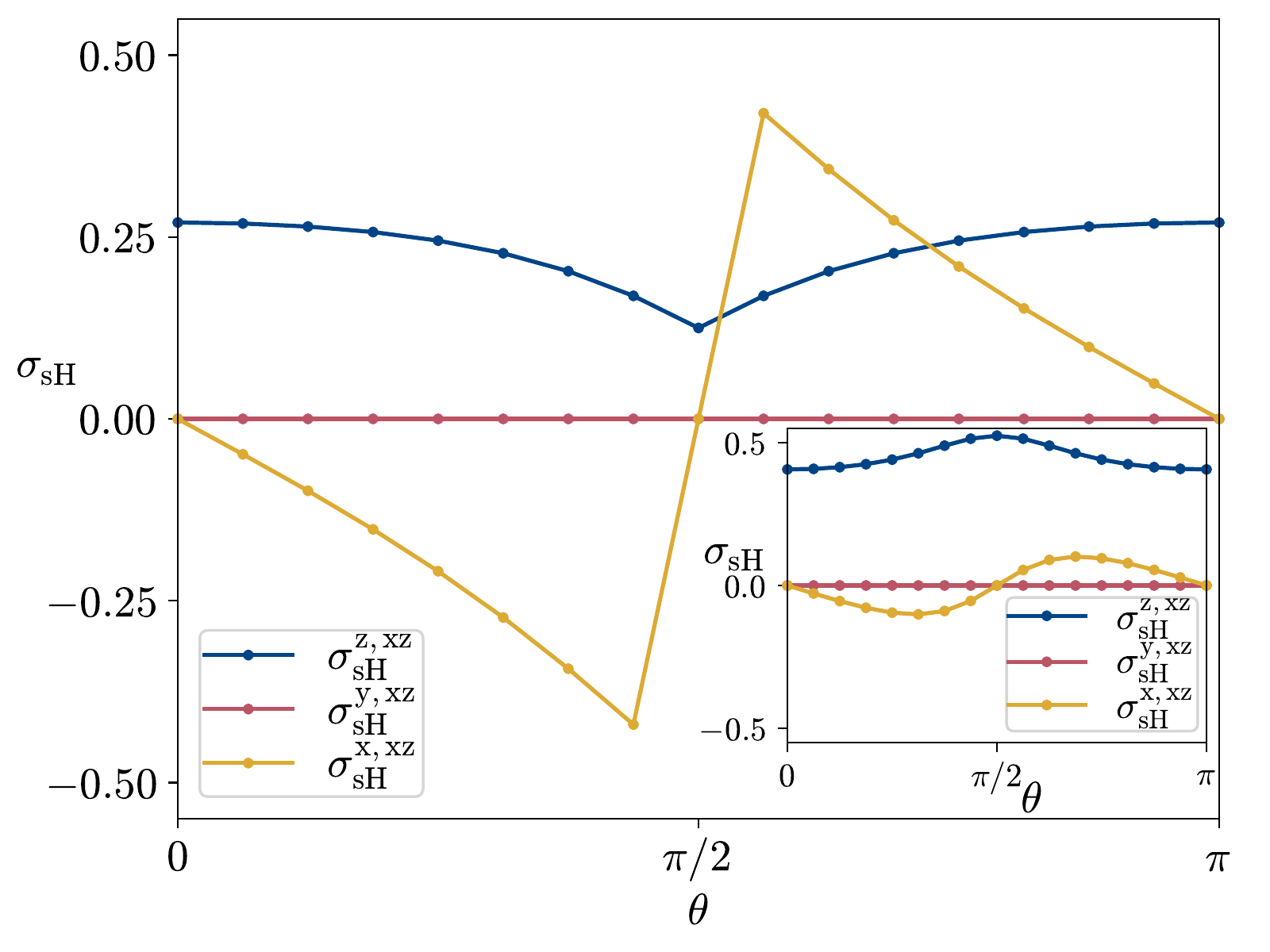}
\caption{AF spin Hall conductivity $\sigma_{\mathrm{sH}}$ for different spin polarizations when the AF spins rotate in the $x$-$z$-plane: $\boldsymbol{n} = (\sin\theta, 0, \cos\theta )$. In this case, the Fermi energy is $E_{\mathrm{F}} = -2t$, the RSOC is $\xi_{\mathrm{R}}=0.1$, and the exchange interaction is $\xi_{\mathrm{sd}} =0.2$. The inset shows similar curves at a larger RSOC: $\xi_{\mathrm{R}}=0.2$. }
	\label{Fig:kuboAFiXZplan}
\end{figure}

First, the AF spins rotate in the $x$-$z$-plane, $\boldsymbol{n} = (\sin\theta, 0, \cos\theta )$. For the $z$-polarization of the spin, the spin Hall conductivity $\sigma^{\mathrm{z,xz}}_{\mathrm{sH}}$ as a function of $\theta$ [Eq. \eqref{eq:appXZplanSigmaZgenereltUttrykk}] is symmetric about $\theta=\pi/2$ (where $\boldsymbol{n}=\hat{x}$), as illustrated in Fig. \ref{Fig:kuboAFiXZplan}. Similar to the case of out-of-plane spins [Sec. \ref{seq:kuboOutofplaneSpins}], the overall amplitude of $\sigma^{\mathrm{z,xz}}_{\mathrm{sH}}$ is mostly determined by $\xi_{\mathrm{sd}}$ and $\xi_{\mathrm{R}}$; varying only $\theta$ results in a small decrease/increase in $\sigma^{\mathrm{z,xz}}_{\mathrm{sH}}$ around $\theta=\pi/2$, as shown in Fig. \ref{Fig:kuboAFiXZplan}. Increasing $\xi_{\mathrm{sd}}$ reduces the amplitude of the spin Hall conductivity. $\sigma^{\mathrm{z,xz}}_{\mathrm{sH}}$ does not vary greatly when changing $E_{\mathrm{F}}$. However, $\sigma^{\mathrm{z,xz}}_{\mathrm{sH}}$ is antisymmetric about $E_{\mathrm{F}}=0$. In Fig. \ref{Fig:kuboAFiXZplan}, we consider the spin Hall conductivities for all three spin polarizations in a system with $E_{\mathrm{F}}=-2t$, $\xi_{\mathrm{sd}}=0.2$, and $\xi_{\mathrm{R}}=0.1$. The inset in Fig. \ref{Fig:kuboAFiXZplan} shows similar curves at a larger RSOC $\xi_{\mathrm{R}}=0.2$. When the AF spins rotate in-plane, the spin Hall conductivity for the $x$-polarization of the spin $\sigma^{\mathrm{x,xz}}_{\mathrm{sH}}$ becomes finite [Eq. \eqref{eq:appXZplanSigmaX}]. As shown in Fig. \ref{Fig:kuboAFiXZplan}, $\sigma^{\mathrm{x,xz}}_{\mathrm{sH}}$ is antisymmetric about $\theta=\pi/2$. The spin Hall conductivity $\sigma^{\mathrm{y,xz}}_{\mathrm{sH}}$ for the $y$-component of the spin vanishes for all $\theta$.

Next, we consider AF spins that rotate in the $y$-$z$-plane, $\boldsymbol{n} = (0, \sin\theta, \cos\theta )$. In this scenario, only the spin Hall conductivity $\sigma^{\mathrm{z,yz}}_{\mathrm{sH}}$ for the $z$-polarization of the spin is finite (results not shown). The spin Hall conductivities $\sigma^{\mathrm{x,yz}}_{\mathrm{sH}}$ and $\sigma^{\mathrm{y,yz}}_{\mathrm{sH}}$, related to the $x$- and $y$-polarizations of the spin, respectively, both vanish. As shown in App. B, from Eq. \eqref{eq:sHallzpolariasjonYZplan}, $\sigma^{\mathrm{z,yz}}_{\mathrm{sH}}$ behaves similarly to $\sigma^{\mathrm{z,xz}}_{\mathrm{sH}}$ when we consider $\sigma^{\mathrm{z,yz}}_{\mathrm{sH}}$ as a function of $\theta$ (results not shown). $\sigma^{\mathrm{z,yz}}_{\mathrm{sH}}$ is symmetric about $\theta=\pi/2$ (where $\boldsymbol{n}=\hat{y}$). The amplitude of $\sigma^{\mathrm{z,yz}}_{\mathrm{sH}}$ is mostly determined by $\xi_{\mathrm{sd}}$ and $\xi_{\mathrm{R}}$; rotating the AF spins in-plane results in moderate changes in $\sigma^{\mathrm{z,yz}}_{\mathrm{sH}}$.

A substantial AF exchange interaction significantly reduces the Berry-phase-induced intrinsic spin Hall conductivity also when the AF spins rotate in-plane.

\section{Disorder and Spin Hall Conductance}\label{section:numerics}

In this section, first, we investigate the electrical characteristics of finite-sized AFs with spin-conserving disorder. We focus on systems where the conducting properties follow Ohm's law. Second, we look at how the spin Hall conductance scales when the size of the disordered AFs changes. Finally, we compare the computed AF spin Hall conductance with the intrinsic Berry phase spin Hall conductivity in bulk systems.

Disorder in the mesoscopic regime is modeled by the onsite elastic potential $\mathcal{V}^{\mathrm{imp}}_{\boldsymbol{r}}\neq 0$ in the Hamiltonian \eqref{eq:hamiltonian_real_space}. The disorder potential is present at all sites in the scattering region. $\mathcal{V}^{\mathrm{imp}}_{\boldsymbol{r}}$ takes a random value uniformly distributed within $-W$ and $W$. The disorder then has zero mean, $\big\langle \mathcal{V}^{\mathrm{imp}}_{\boldsymbol{r}} \big\rangle=0$, and the variance is $\big\langle (\mathcal{V}^{\mathrm{imp}}_{\boldsymbol{r}})^{2} \big\rangle - \big\langle \mathcal{V}^{\mathrm{imp}}_{\boldsymbol{r}} \big\rangle^{2} = (1/3)W^{2}$. We calculate the average of the transport properties by considering many disorder configurations.

\subsection{Ohmic Regime}\label{Sec:OmhicRegion}

We now introduce disorder ($\mathcal{V}^{\mathrm{imp}}_{\boldsymbol{r}}\neq 0$) in AFs in a system that is similar to Fig. \ref{Fig:fourterminalAF}, as discussed in Sec. \ref{section:theory}. In this section, we change the geometry of the AF by considering a scattering region of a rectangular shape [instead of the square shape of area $(Na)^{2}$]. We only consider AFs of rectangular shapes in this part (Sec. \ref{Sec:OmhicRegion}).

The length of the AF conductor (scattering region) is $N_{\mathrm{x}}a$, where $N_{\mathrm{x}}$ is the number of sites in the $x$-direction. The width of the conductor is $N_{\mathrm{y}}$a, where $N_{\mathrm{y}}$ is the number of sites in the $y$-direction. Four leads are attached to the sides of the scattering region, and the widths of the leads are determined by $N_{\mathrm{x}}a$ (top/bottom leads) and $N_{\mathrm{y}}a$ (left/right leads). Otherwise, we use the same boundary conditions as in Sec. \ref{section:theory}.

To estimate when the AFs are in the ohmic regime, we define the {\it effective two-terminal conductance} $g^{\mathrm{eff}}_{\mathrm{c}}$ as
\begin{equation}\label{eq:electricConductanceDef}
g^{\mathrm{eff}}_{\mathrm{c}} = \frac{1}{2}(g_{01} + g_{10}) \, ,
\end{equation}
which is the average transmission for an electron originating in the left (right) lead scattering into the right (left) lead. We are interested in how the electrical resistance of the AF scales as the length of the conductor varies. For systems in two spatial dimensions, the effective charge conductivity $\sigma^{\mathrm{eff}}_{\mathrm{c}}$ can be suitably defined as
\begin{equation}\label{eq:elecConductivityDef}
\sigma^{\mathrm{eff}}_{\mathrm{c}} = g^{\mathrm{eff}}_{\mathrm{c}} \frac{ N_{\mathrm{x}} }{ N_{\mathrm{y}} } \,
\end{equation}
based on the effective two-terminal conductance $g^{\mathrm{eff}}_{\mathrm{c}}$ from Eq. \eqref{eq:electricConductanceDef}. When $\sigma^{\mathrm{eff}}_{\mathrm{c}}$ is {\it independent} of the system length $N_{\mathrm{x}}a$, the AF conductor is in the ohmic regime, analogous to Ohm's law for a conductor where the charge current is driven by an electric field $E=\Delta V/(N_x a)$ in terms of the potential difference $\Delta V$ between two terminals.
\begin{figure}[h]
	\centering
	\includegraphics[width=1.0\linewidth]{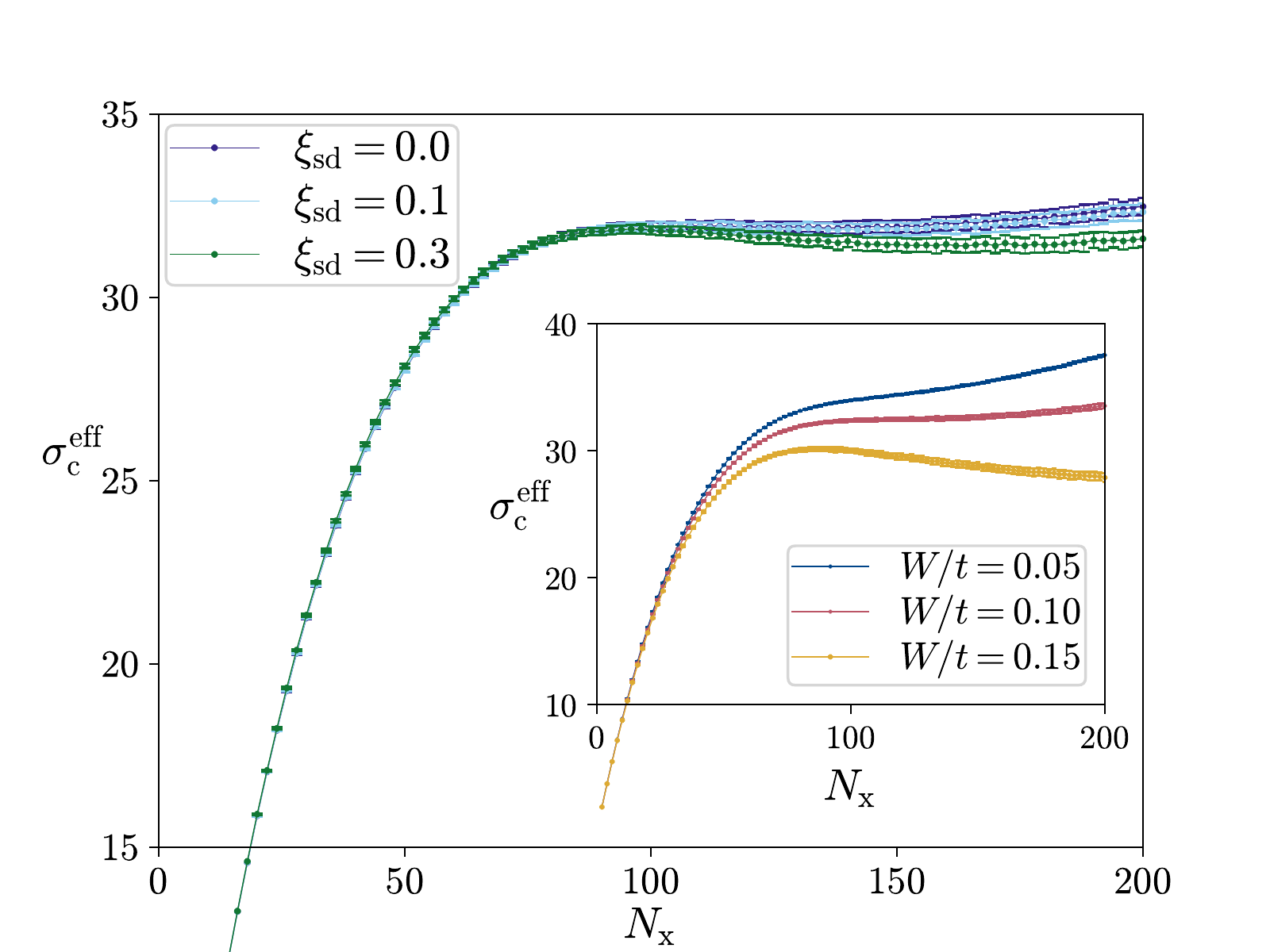}
\caption{Effective charge conductivity $\sigma^{\mathrm{eff}}_{\mathrm{c}}$ as a function of length $N_{\mathrm{x}}$, with $N_{\mathrm{x}}$ even. The different curves are for AFs with out-of-plane spins ($\boldsymbol{n}=\hat{z}$) with different exchange couplings $\xi_{\mathrm{sd}}$ but otherwise the same parameters. The points show $\sigma^{\mathrm{eff}}_{\mathrm{c}}$ as an average over disorder realizations. The error bars show the standard deviation. Each point is averaged over 100 disorder configurations with disorder strength $W=0.11t$. The width of the AF is kept constant, $N_{\mathrm{y}}=100$. The Fermi energy is $E_{\mathrm{F}}=-2t$, and the RSOC is $\xi_{\mathrm{R}} = 0.1$. The inset shows similar curves in the NM limit ($\xi_{\mathrm{sd}}=0$), where the different curves correspond to different disorder strengths $W$ but otherwise the same parameters. }
	\label{Fig:electricCondsLengthX}
\end{figure}

We now present the typical results for the effective charge conductivity $\sigma^{\mathrm{eff}}_{\mathrm{c}}$ of the AFs (and NMs). In Fig.\ \ref{Fig:electricCondsLengthX}, we plot $\sigma^{\mathrm{eff}}_{\mathrm{c}}$ as a function of even $N_{\mathrm{x}}$ for AFs with out-of-plane spins ($\boldsymbol{n}=\hat{z}$). The different curves correspond to increasing exchange coupling $\xi_{\mathrm{sd}}$. In evaluating $\sigma^{\mathrm{eff}}_{\mathrm{c}}$, we average over 100 disorder configurations with disorder strength $W=0.11t$. The error bars show the standard deviation. The other parameters used in Fig.\ \ref{Fig:electricCondsLengthX} are $E_{\mathrm{F}}=-2t$ and $\xi_{\mathrm{R}}=0.1$. The AF width is fixed at $N_{\mathrm{y}}a = 100a$. In Fig. \ref{Fig:electricCondsLengthX}, $\sigma^{\mathrm{eff}}_{\mathrm{c}}$ saturates in the region of $N_{\mathrm{x}}$ between $100$ and $200$ for several exchange couplings $\xi_{\mathrm{sd}}$. Increasing $\xi_{\mathrm{sd}}$ to an amplitude of $|\xi_{\mathrm{sd}}| \sim 1$
does not significantly alter the regions where $\sigma^{\mathrm{eff}}_{\mathrm{c}}$ is independent of the system length. However, increasing $|\xi_{\mathrm{sd}}|$ to above 1 (similar to the cases discussed in Sec. \ref{sec_sub_AFs_ballistic}) increases the electrical resistance in the AF such that the electrons become localized. The slope of $\sigma^{\mathrm{eff}}_{\mathrm{c}}$ increases with the RSOC $\xi_{\mathrm{R}}$ because spin flip allows more conducting paths for the electron flow. The inset in Fig. \ref{Fig:electricCondsLengthX} illustrates how the effective charge conductivity $\sigma^{\mathrm{eff}}_{\mathrm{c}}$ for an NM changes when the disorder strength $W$ is varied around $W\sim 0.1t$. An increase in the disorder strength $W$ results in a $\sigma^{\mathrm{eff}}_{\mathrm{c}}$ that rapidly decreases as a function of conductor length $N_{\mathrm{x}}a$. Similar results are found for the relevant AFs when varying $W$.

Based on our considerations of the effective charge conductivity $\sigma^{\mathrm{eff}}_{\mathrm{c}}$, the AFs are close to the ohmic regime when the disorder strength is approximately $W\approx0.1t$, the AF exchange couplings obey $|\xi_{\mathrm{sd}}| \lesssim 1 $, and the RSOC obeys $|2\xi_{\mathrm{R}}|\lesssim 1$. Similar results are found at different Fermi energies $E_{\mathrm{F}}$ (Fig. \ref{Fig:electricCondsLengthX} shows the results at $E_{\mathrm{F}}=-2t$) for the system sizes we considered.

\subsection{Spin Hall Conductance and Ohm's Law}\label{sec:shConductanceAndOhmsLawLengths}

We now investigate how the spin Hall conductance $g_{\mathrm{sH}}$ in disordered AFs scales as the system size increases. The disorder strength $W$ and the other system parameters of the AFs are chosen such that the electrical properties are in the ohmic regime (see Sec. \ref{Sec:OmhicRegion}).

A meaningful comparison between the spin Hall conductance $g_{\mathrm{sH}}$ in mesoscopic systems and the intrinsic spin Hall conductivity $\sigma_{\mathrm{sH}}$ in bulk systems requires that the spin Hall conductance $g_{\mathrm{sH}}$ be independent of the system size. Ohm's law for charge currents relates the electrical conductance to the electrical conductivity, similar to Eq. \eqref{eq:elecConductivityDef}. Similarly, we can envision an Ohm's law for spin currents, which relates $g_{\mathrm{sH}}$ to $\sigma_{\mathrm{sH}}$. We focus on AFs with a square shape of area $(Na)^{2}$. For such systems, when $g_{\mathrm{sH}}$ is {\it constant} as the AF width $Na$ varies, we can directly compare the spin Hall conductance to the spin Hall conductivity, i.e., $g_{\mathrm{sH}} = \sigma_{\mathrm{sH}}$.
\begin{figure}[h]
	\centering
	\includegraphics[width=1.0\linewidth]{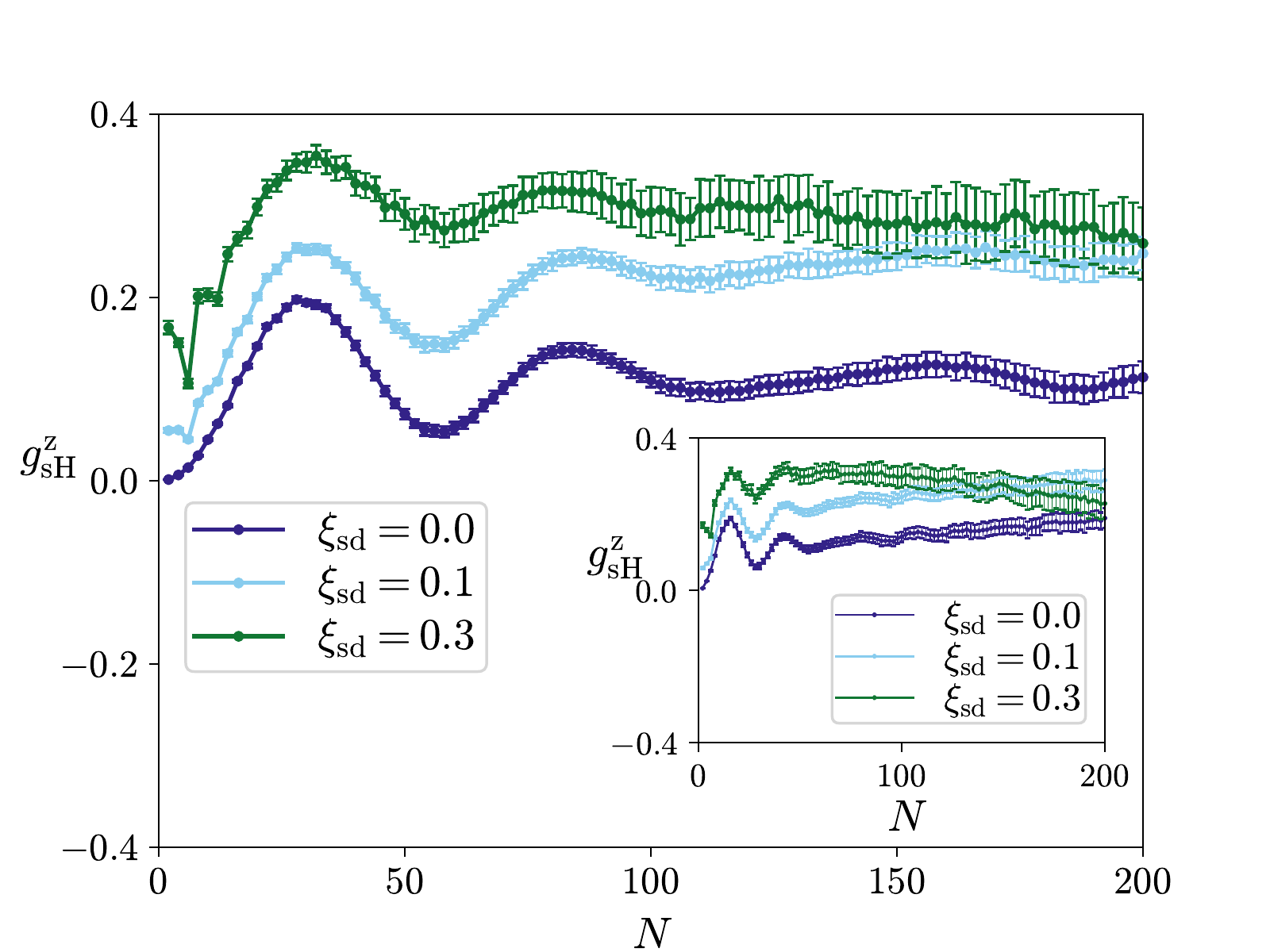}
\caption{Spin Hall conductance $g^{\mathrm{z}}_{\mathrm{sH}}$ as a function of size $N$, with $N$ even. The curves show AFs with out-of-plane spins ($\boldsymbol{n}=\hat{z}$) with different exchange couplings $\xi_{\mathrm{sd}}$. Here, $g^{\mathrm{z}}_{\mathrm{sH}}$ is averaged over 100 disorder realizations, with disorder strength $W=0.11t$. The error bars show the standard deviation. The RSOC is $\xi_{\mathrm{R}} = 0.1$, and the Fermi energy is $E_{\mathrm{F}}=-2t$. The inset shows similar results at a larger RSOC, $\xi_{\mathrm{R}} = 0.2$. }
	\label{Fig:AFshLenghtsMedDisorder}
\end{figure}

Here, we present results for $g^{\mathrm{z}}_{\mathrm{sH}}$ as a function of size $N$. We consider AFs with $N$ up to 200. In Fig. \ref{Fig:AFshLenghtsMedDisorder}, the spin Hall conductance $g^{\mathrm{z}}_{\mathrm{sH}}$ as a function of even $N$ is shown for AFs with out-of-plane spins $(\boldsymbol{n}=\hat{z})$ for several exchange couplings $\xi_{\mathrm{sd}}$. The values of $g^{\mathrm{z}}_{\mathrm{sH}}$ in Fig. \ref{Fig:AFshLenghtsMedDisorder} are averaged over 100 disorder realizations at disorder strength $W=0.11t$, and the error bars show the standard deviation. The common system parameters for the curves in Fig. \ref{Fig:AFshLenghtsMedDisorder} are an RSOC of $\xi_{\mathrm{R}}=0.1$ and a Fermi energy of $E_{\mathrm{F}}=-2t$. The inset of Fig. \ref{Fig:AFshLenghtsMedDisorder} shows similar results for a larger RSOC of $\xi_{\mathrm{R}} = 0.2$. As shown in Fig. \ref{Fig:AFshLenghtsMedDisorder}, most of the curves for $g^{\mathrm{z}}_{\mathrm{sH}}$ do not vary greatly when the length varies between $Na=100a$ and $Na=200a$.

We have focused on AFs in the ohmic regime where $W\sim 0.1t$, while the RSOC and the exchange interaction are of intermediate strengths. The general trend in this regime is that $g^{\mathrm{z}}_{\mathrm{sH}}$ does not vary greatly as a function of width $Na$ for $N\sim 100$ and greater, similar to the results illustrated in Fig. \ref{Fig:AFshLenghtsMedDisorder}. $g^{\mathrm{z}}_{\mathrm{sH}}$ as a function of $N$ varies more when the RSOC and exchange interaction are larger. The standard deviation of $g^{\mathrm{z}}_{\mathrm{sH}}$ increases when $W$, $\xi_{\mathrm{R}}$, and $\xi_{\mathrm{sd}}$ increase. Typically, when $|2\xi_{\mathrm{R}}|$ and/or $|\xi_{\mathrm{sd}}|$ are large (on the order of $1$), the spin Hall conductance $g^{\mathrm{z}}_{\mathrm{sH}}$ depends on the system size.

In general, apart from the smallest systems ($N\lesssim100$), $g^{\mathrm{z}}_{\mathrm{sH}}$ does not vary greatly with size for $W\approx 0.1t$, exchange coupling $|\xi_{\mathrm{sd}}| \lesssim 0.5 $, and RSOC $|2\xi_{\mathrm{R}}|\lesssim 0.5$. Within this regime, the spin currents in the top/bottom leads are pure spin currents because the charge currents induced in the transverse leads are very small compared to the spin currents. The results are similar at different Fermi energies $E_{\mathrm{F}}$ (Fig. \ref{Fig:AFshLenghtsMedDisorder} shows the results at $E_{\mathrm{F}}=-2t$). We take these results into consideration when we compare the spin Hall conductance $g_{\mathrm{sH}}$ to the spin Hall conductivity $\sigma_{\mathrm{sH}}$ in Secs. \ref{sec:disorderdNMs} and \ref{sec:AFsWithDisorder} for NMs and AFs, respectively.

\subsection{Disordered Normal Metals}\label{sec:disorderdNMs}

Here, we show spin Hall conductance $g^{\mathrm{z}}_{\mathrm{sH}}$ results for NMs ($\xi_{\mathrm{sd}}=0$) with disorder. We consider NMs in the ohmic regime ($W\approx 0.1t$) for an RSOC $\xi_{\mathrm{R}}$ of intermediate strength, where $g^{\mathrm{z}}_{\mathrm{sH}}$ is approximately independent of size (see Sec. \ref{sec:shConductanceAndOhmsLawLengths}). $g^{\mathrm{z}}_{\mathrm{sH}}$ attains a constant value independent of the RSOC and the Fermi energy ($g^{\mathrm{z}}_{\mathrm{sH}}$ is antisymmetric about $E_{\mathrm{F}}=0$). The spin Hall conductances related to the $x$- and $y$-polarizations of the spin vanish, similar to in ballistic NMs.
\begin{figure}[h]
	\centering
	\includegraphics[width=1.0\linewidth]{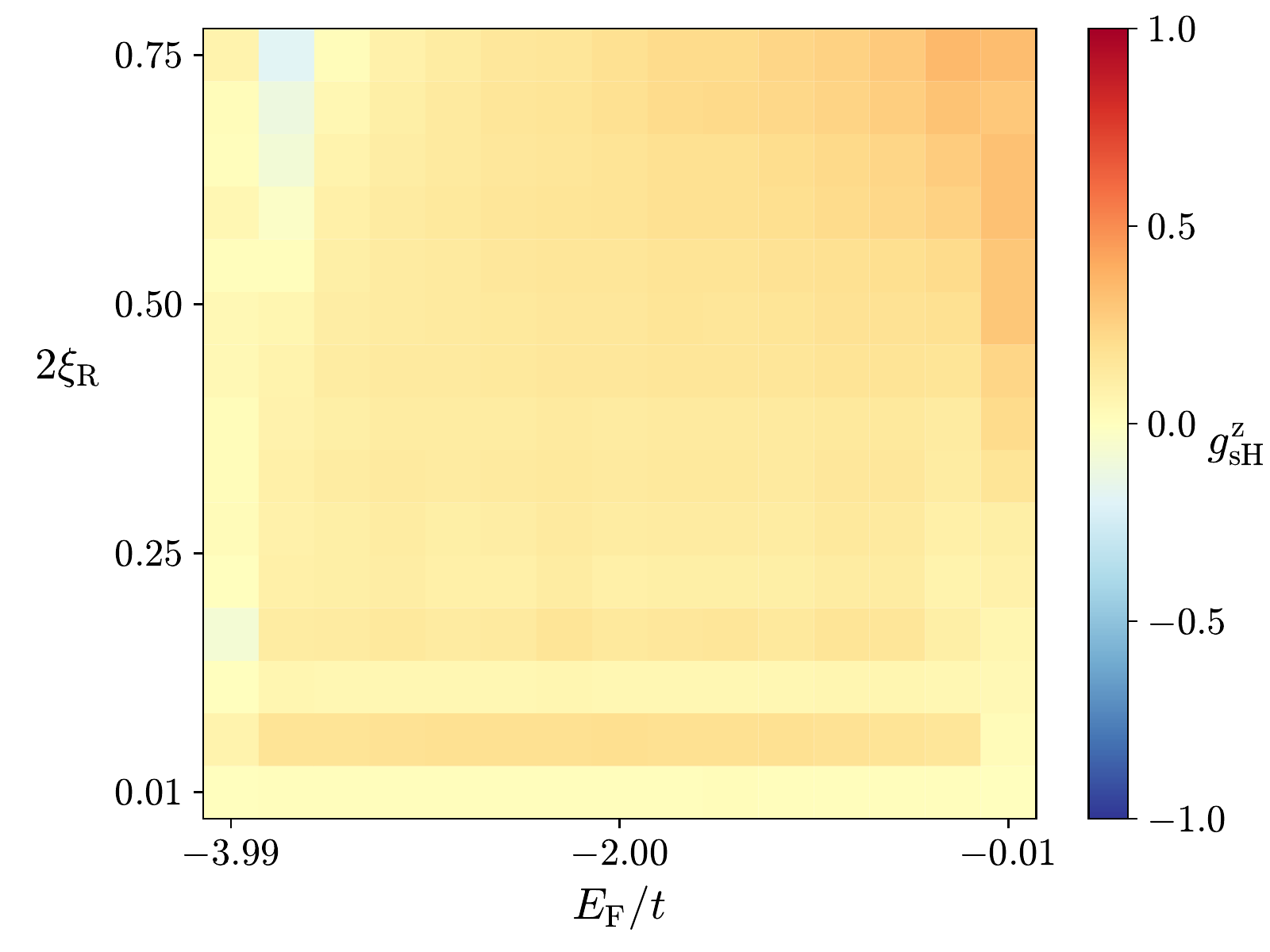}
\caption{Spin Hall conductance $g^{\mathrm{z}}_{\mathrm{sH}}$ in the NM limit ($\xi_{\mathrm{sd}} = 0$) as a function of Fermi energy $E_{\mathrm{F}}$ and RSOC $2\xi_{\mathrm{R}}$. Here, each square represents one value of $g^{\mathrm{z}}_{\mathrm{sH}}$ averaged over 100 disorder configurations. The system size is $N^{2}=100^{2}$, while the disorder strength is $W=0.11t$. }
	\label{Fig:2dNMlimit}
\end{figure}

In Fig. \ref{Fig:2dNMlimit}, we plot the spin Hall conductance $g^{\mathrm{z}}_{\mathrm{sH}}$ for an NM as a function of RSOC $2\xi_{\mathrm{R}}$ and Fermi energy $E_{\mathrm{F}}$ for a system of size $N^{2}=100^{2}$. We average $g^{\mathrm{z}}_{\mathrm{sH}}$ over 100 disorder configurations at disorder strength $W=0.11t$. As illustrated in Fig. \ref{Fig:2dNMlimit}, the spin Hall conductance $g^{\mathrm{z}}_{\mathrm{sH}}$ is approximately constant when varying the Fermi energy and the RSOC, and the average value of all points in Fig. \ref{Fig:2dNMlimit} is $g^{\mathrm{z}}_{\mathrm{sH}} \approx 0.13$. In Fig. \ref{Fig:2dNMlimit}, for the few points closest to $E_{\mathrm{F}}\approx -4t$ and $E_{\mathrm{F}}\approx0$, the standard deviations $\sim 0.1$ are on the order of $g^{\mathrm{z}}_{\mathrm{sH}}$; otherwise, the standard deviations $\sim 0.01$ are much smaller than $g^{\mathrm{z}}_{\mathrm{sH}}$.

For NMs, we can compare the spin Hall conductance $g^{\mathrm{z}}_{\mathrm{sH}}$ in mesoscopic systems with the Berry-phase-induced intrinsic spin Hall conductivity $\sigma^{\mathrm{z}}_{\mathrm{sH}}$ in bulk systems. Both $g^{\mathrm{z}}_{\mathrm{sH}}$ and $\sigma^{\mathrm{z}}_{\mathrm{sH}}$ take a constant value, independent of the RSOC $\xi_{\mathrm{R}}$ and the Fermi energy $E_{\mathrm{F}}$, as illustrated in Figs. \ref{Fig:2dNMlimit} and \ref{Fig:kuboNormalMetall}, respectively. When the Fermi energy obeys $-4t < E_{\mathrm{F}} <0$, the values of $g^{\mathrm{z}}_{\mathrm{sH}} \sim 0.1$ and $\sigma^{\mathrm{z}}_{\mathrm{sH}}\approx 0.5$ are on the same order but somewhat different. We note here that for NMs with RSOC in the continuum model, the contribution to the spin Hall conductivity $\sigma^{\mathrm{z}}_{\mathrm{sH}}$ from the Berry phase is 0.5, while vertex corrections yield a vanishing $\sigma^{\mathrm{z}}_{\mathrm{sH}}$ \cite{sinovaUniversalPRL2004,inouePRB2004jul,mishchenkoPRL2004nov,olegPRB2005juni,olgaPRB2005jun,krotkovPRB2006,SINOVA2006214}. In our tight-binding model, $\sigma^{\mathrm{z}}_{\mathrm{sH}}\approx 0.5$ coincides with the result in the continuum limit. However, $g^{\mathrm{z}}_{\mathrm{sH}} \sim 0.1$ is between the results for the continuum model with and without vertex corrections.

\subsection{Antiferromagnets with Disorder}\label{sec:AFsWithDisorder}

In this part, we present results for the spin Hall conductance $g^{\mathrm{z}}_{\mathrm{sH}}$ in AFs with spin-conserving disorder. We consider AFs in a regime where the conductive properties behave similar to Ohm's law, where $g^{\mathrm{z}}_{\mathrm{sH}}$ does not change greatly for different system sizes, as discussed in Secs. \ref{Sec:OmhicRegion} and \ref{sec:shConductanceAndOhmsLawLengths}. Within this regime, we compare the mesoscopic spin Hall conductance $g^{\mathrm{z}}_{\mathrm{sH}}$ in AFs to the intrinsic spin Hall conductivity $\sigma^{\mathrm{z}}_{\mathrm{sH}}$. In the following, we focus on $g^{\mathrm{z}}_{\mathrm{sH}}$, related to the $z$-component of the spin, in AFs with out-of-plane localized spins.
\begin{figure}[h]
	\centering
	\includegraphics[width=1.0\linewidth]{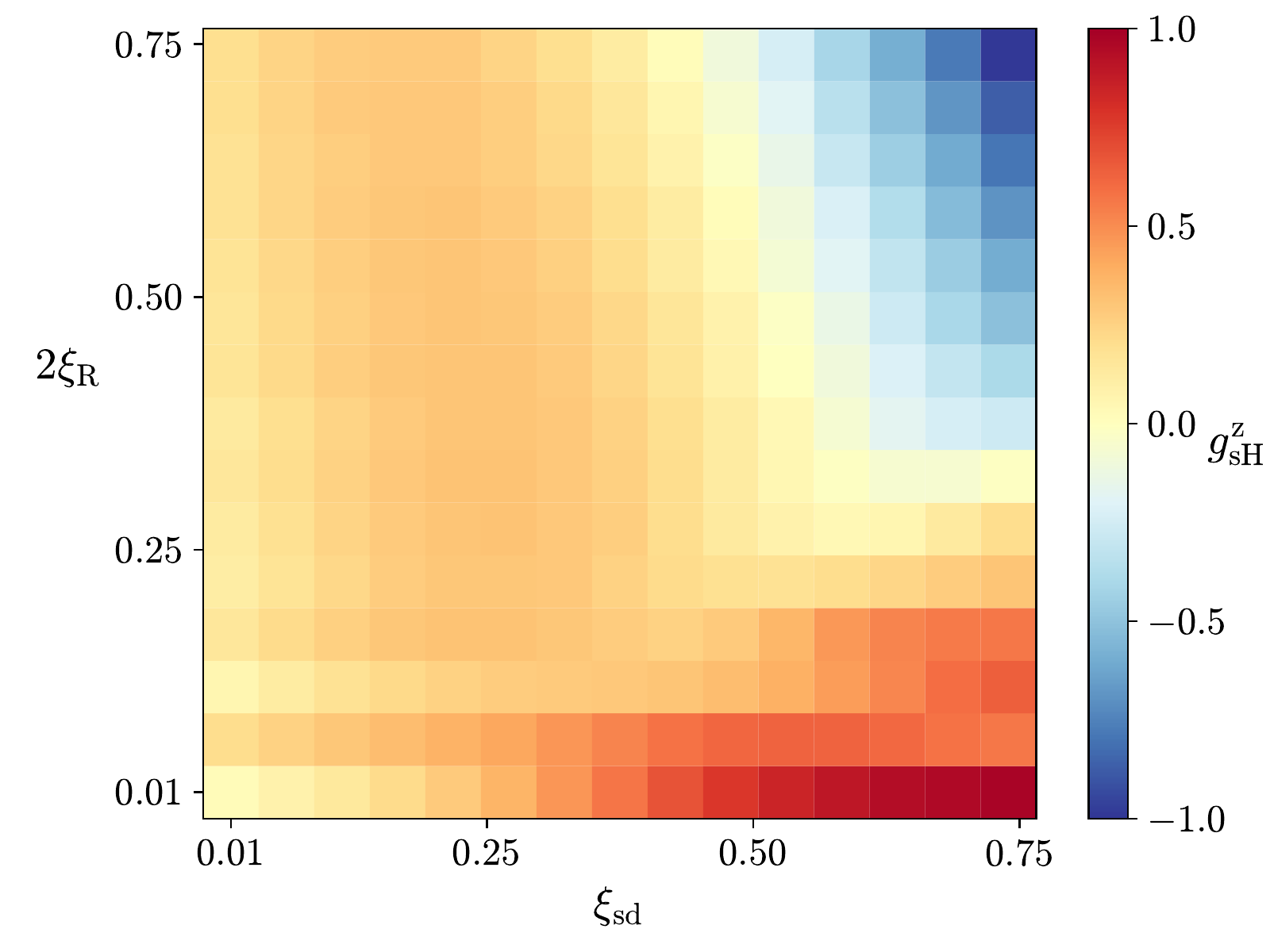}
\caption{Spin Hall conductance $g^{\mathrm{z}}_{\mathrm{sH}}$ for an AF at Fermi energy $E_{\mathrm{F}}=-2t$, plotted as a function of exchange interaction $\xi_{\mathrm{sd}}$ and RSOC $2\xi_{\mathrm{R}}$. Each value of $g^{\mathrm{z}}_{\mathrm{sH}}$ is calculated as an average over 100 disorder configurations, where the disorder strength is $W=0.11t$. The system size is $N^{2}=100^{2}$. }
	\label{Fig:AF2dEfMin2}
\end{figure}

The spin Hall conductance $g^{\mathrm{z}}_{\mathrm{sH}}$ in AFs with disorder ($W\sim 0.1t$) shows quite different behavior at different Fermi energies. In many cases, $g^{\mathrm{z}}_{\mathrm{sH}}$ varies considerably as a function of the RSOC $\xi_{\mathrm{R}}$ and exchange interaction $\xi_{\mathrm{sd}}$ in the AF.
\begin{figure}[h]
	\centering
	\includegraphics[width=1.0\linewidth]{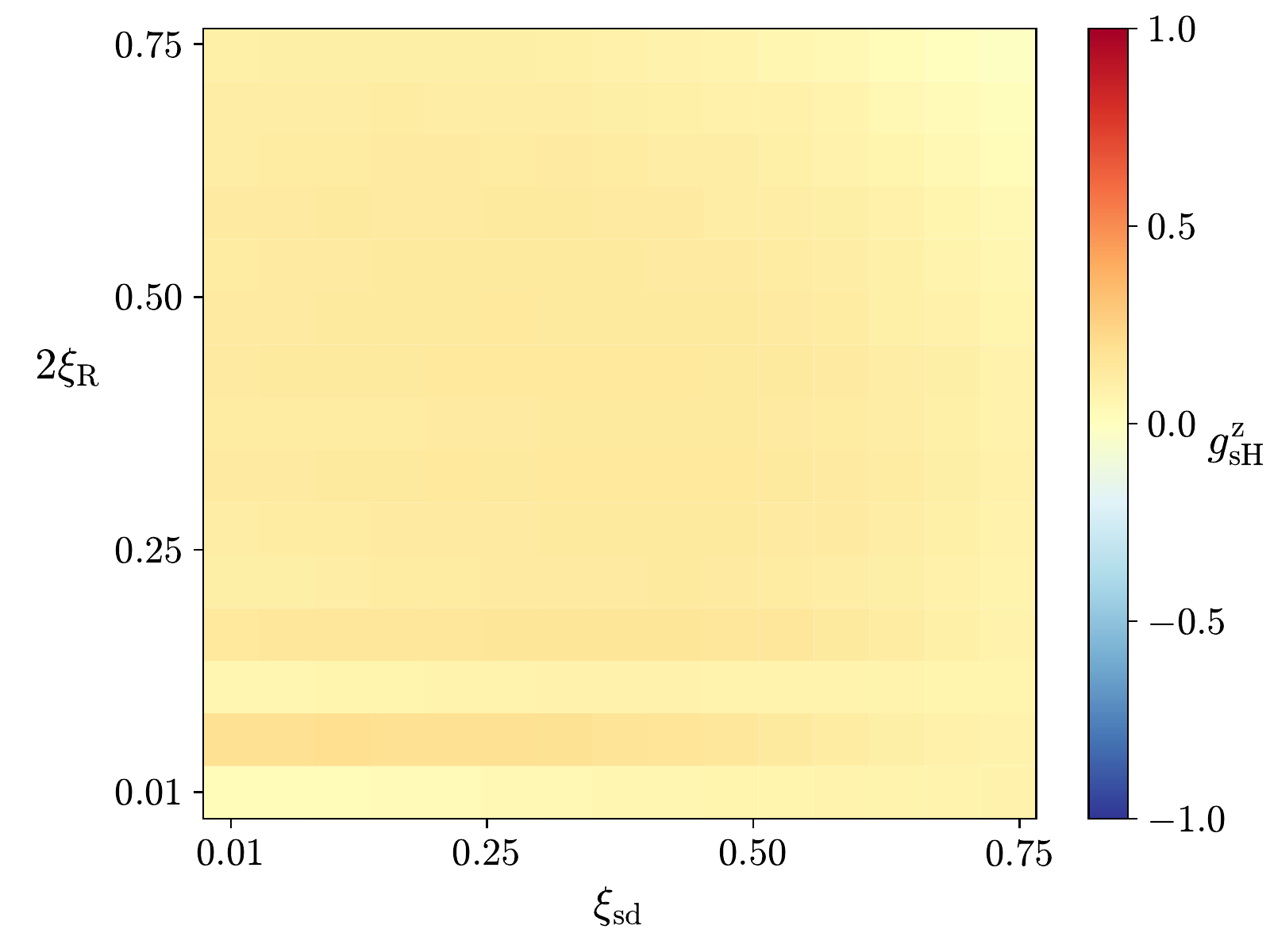}
\caption{Spin Hall conductance $g^{\mathrm{z}}_{\mathrm{sH}}$ for an AF with Fermi energy $E_{\mathrm{F}}=-3t$. The RSOC is determined by $2\xi_{\mathrm{R}}$, while $\xi_{\mathrm{sd}}$ is the strength of the exchange interaction. The size of the AF is $N^{2}=100^{2}$. Each square represents $g^{\mathrm{z}}_{\mathrm{sH}}$ as an average over 100 disorder realizations at disorder strength $W=0.11t$. }
	\label{Fig:AF2dEfMin3}
\end{figure}

Figs. \ref{Fig:AF2dEfMin2} and \ref{Fig:AF2dEfMin3} show the results for $g^{\mathrm{z}}_{\mathrm{sH}}$ in AFs at different Fermi energies: $E_{\mathrm{F}} =-2t$ and $E_{\mathrm{F}}=-3t$, respectively. Apart from the different values of $E_{\mathrm{F}}$, the AFs in Figs. \ref{Fig:AF2dEfMin2} and \ref{Fig:AF2dEfMin3} have the same system parameters. The system size is $N^{2}=100^{2}$. $g^{\mathrm{z}}_{\mathrm{sH}}$ is averaged over 100 disorder configurations, with disorder strength $W=0.11t$. At $E_{\mathrm{F}}=-3t$, the spin Hall conductance $g^{\mathrm{z}}_{\mathrm{sH}}\sim 0.1$ does not vary greatly when varying the RSOC $\xi_{\mathrm{R}}$ and/or the exchange interaction $\xi_{\mathrm{sd}}$, as shown in Fig. \ref{Fig:AF2dEfMin3}. However, at Fermi energy $E_{\mathrm{F}}=-2t$, Fig. \ref{Fig:AF2dEfMin2} illustrates that $g^{\mathrm{z}}_{\mathrm{sH}}$ varies considerably between $1.0$ and $-1.0$ when changing $\xi_{\mathrm{R}}$ and $\xi_{\mathrm{sd}}$. For intermediate RSOC $\xi_{\mathrm{R}}$, Fig. \ref{Fig:AF2dEfMin2} illustrates that, in certain regimes, increasing the exchange interaction $\xi_{\mathrm{sd}}$ in AFs can increase the spin Hall conductance $g^{\mathrm{z}}_{\mathrm{sH}}$, similar to the results in Sec. \ref{sec_sub_AFs_ballistic}. In Figs. \ref{Fig:AF2dEfMin2} and \ref{Fig:AF2dEfMin3}, the typical standard deviations in $g^{\mathrm{z}}_{\mathrm{sH}}$ are greatest when $\xi_{\mathrm{R}}$ and/or $\xi_{\mathrm{sd}}$ are large, on the order of $0.02$.

The spin Hall conductance $g^{\mathrm{z}}_{\mathrm{sH}}$ is antisymmetric with respect to $E_{\mathrm{F}} = 0$, which is due to particle-hole symmetry \cite{nikolicPRBaug2005}. The energy dispersion of the propagating modes are symmetric with respect to $E_{\mathrm{F}} = 0$. Changing the sign of the Fermi energy ($E_{\mathrm{F}}\rightarrow -E_{\mathrm{F}}$) transforms electron-like carriers into hole-like carriers, and vice versa, such that $g^{\mathrm{z}}_{\mathrm{sH}}$ changes sign. The magnitude of the spin Hall conductance depends on the ratio between the number of available propagating modes to the transverse size $N$. This ratio changes as a function of Fermi energy. Close to the band bottom ($E_{\mathrm{F}}$ close to $-4t$) there are less available modes per transverse site $N$ that contribute to the spin Hall conductance. When $E_{\mathrm{F}}$ is close to 0, there are more modes relative to the transverse size that can contribute to the spin transport. For instance, this indicates why the overall amplitude of the spin Hall conductance is smaller when the Fermi energy is $-3t$ as compared to the results at $E_{\mathrm{F}} = -2t$, as shown in Figs. \ref{Fig:AF2dEfMin2} and \ref{Fig:AF2dEfMin3}.

Now, we compare the results for the spin Hall conductance $g^{\mathrm{z}}_{\mathrm{sH}}$ in AFs with disorder ($W\sim 0.1t$) to those for the intrinsic spin Hall conductivity $\sigma^{\mathrm{z}}_{\mathrm{sH}}$ induced by the Berry phase term in the Kubo formula (Sec. \ref{sec:sub:Kubo}). As functions of Fermi energy, both $g^{\mathrm{z}}_{\mathrm{sH}}$ and $\sigma^{\mathrm{z}}_{\mathrm{sH}}$ are antisymmetric about $E_{\mathrm{F}}=0$ (we discuss $-4t<E_{\mathrm{F}}<0$ in the following). In AFs, the results for $g^{\mathrm{z}}_{\mathrm{sH}}$ are quite different from those for $\sigma^{\mathrm{z}}_{\mathrm{sH}}$. The spin Hall conductance $g^{\mathrm{z}}_{\mathrm{sH}}$ varies considerably for different Fermi energies $E_{\mathrm{F}}$, while the Berry phase contribution to the spin Hall conductivity $\sigma^{\mathrm{z}}_{\mathrm{sH}}$ is similar at different $E_{\mathrm{F}}$. In some regimes (e.g., see Fig. \ref{Fig:AF2dEfMin2}), an increasing exchange interaction and/or varying RSOC $\xi_{\mathrm{R}}$ can enhance the amplitude of the spin Hall conductance $g^{\mathrm{z}}_{\mathrm{sH}}$. In contrast, the qualitative picture of the spin Hall conductivity $\sigma^{\mathrm{z}}_{\mathrm{sH}}$ is that increasing the exchange interactions $\xi_{\mathrm{sd}}$ in AFs reduces the SHE.


\section{Conclusions}\label{section:conclusion}

We considered the spin Hall effect in antiferromagnets. As a description, we use a tight-binding model with Rashba spin-orbit coupling and a staggered on-site exchange field. We also study how on-site spin-independent disorder governs transport properties.

The mesoscopic spin Hall conductance $g^{\mathrm{z}}_{\mathrm{sH}}$ varies considerably and is sensitive to the parameters of the system, such as the Fermi energy, the exchange interaction, and the spin-orbit interaction. In some regimes, increasing the exchange interactions increases the spin Hall conductance $g^{\mathrm{z}}_{\mathrm{sH}}$. The spin Hall effect in antiferromagnets may, therefore, be more substantial than the corresponding effect in the normal state. This is somewhat surprising to us since, naively, the staggered exchange coupling competes with the spin-orbit coupling. A larger exchange interaction align the itinerant spins more strongly to the Neel field. This alignment could reduce the impact of the spin-orbit coupling. Indeed, this happens for very large strengths of the exchange interaction, but for more moderate values we find that the intricate interplay with the spin-orbit coupling produces the opposite effect.

When the disorder, the spin-orbit coupling, and the exchange interaction are of intermediate strengths, we identify a diffusive regime where we can make a qualitative comparison between the mesoscopic spin Hall conductance $g^{\mathrm{z}}_{\mathrm{sH}}$ and the Berry-phase contribution to the spin Hall conductivity $\sigma^{\mathrm{z}}_{\mathrm{sH}}$. Compared to our mesoscopic antiferromagnet, the spin Hall conductivity $\sigma^{\mathrm{z}}_{\mathrm{sH}}$ calculated from the Berry-phase contribution in the Kubo formula behaves differently than the results from the exact numerical diagonalization of the spin Hall conductance. For instance, an increasing exchange interaction in antiferromagnets significantly reduces the Berry-phase contribution to the spin Hall conductivity, opposite to the exact numerical results.

As in other systems, our results demonstrate that computing the spin Hall effect in antiferromagnets from the Berry-phase-induced spin Hall conductivity in the Kubo formula is insufficient and neither quantitatively nor qualitatively reproduces the exact numerical results for a disordered system.


\begin{acknowledgements}
This work received funding from the European Research Council via Advanced Grant No. 669442 "Insulatronics" as well as the Research Council of Norway via Grant No. 239926 "Super Insulator Spintronics" and through its Centre of Excellence funding scheme "QuSpin" Grant No. 262633.
\end{acknowledgements}


\appendix

\section{Spin and Charge Currents}\label{appendix:TBcurrents}

The time evolution of an operator in the Heisenberg picture $\hat{\mathcal{O}}(t) = e^{i\hat{H}t/\hbar}\hat{\mathcal{O}}e^{-i\hat{H}t/\hbar}$ is determined from the commutator $\hbar\partial_{t}\hat{\mathcal{O}}(t)=i\big[\hat{H}, \hat{\mathcal{O}}(t)\big]$, with $\hat{H}$ being the Hamiltonian. For brevity, we omit the Heisenberg time dependence in the following.

The itinerant charge density operator at position $\boldsymbol{r}$ is $\hat{\rho}_{\boldsymbol{r}} = -e\hat{c}^{\dagger}_{\boldsymbol{r}}\hat{c}_{\boldsymbol{r}}$. From the time rate of change of the charge density operator and using $\hat{H}$ from Eq. \eqref{eq:hamiltonian_real_space}, the continuity equation for the charge current is
\begin{equation}\label{eq:continuityEqChargeGeneral}
0 =  \frac{\partial}{\partial t}\hat{\rho}_{\boldsymbol{r}}  + \sum_{\boldsymbol{\delta}=\pm\boldsymbol{\delta}_{\mathrm{x}},\pm\boldsymbol{\delta}_{\mathrm{y}}} \hat{j}^{\mathrm{c}}_{\boldsymbol{r},\boldsymbol{\delta}}  \, .
\end{equation}
In Eq. \eqref{eq:continuityEqChargeGeneral}, the operator $\hat{j}^{\mathrm{c}}_{\boldsymbol{r},\boldsymbol{\delta}}$ denotes the charge current between lattice sites $\boldsymbol{r}$ and $\boldsymbol{r}+\boldsymbol{\delta}$. The charge current operator is separated into the two terms
\begin{equation}\label{eq:chargeCurrentsTwoTerms}
\hat{j}^{\mathrm{c}}_{\boldsymbol{r},\boldsymbol{\delta}} = \hat{j}^{\mathrm{c,}0}_{\boldsymbol{r},\boldsymbol{\delta}} + \hat{j}^{\mathrm{c,a}}_{\boldsymbol{r},\boldsymbol{\delta}} \, ,
\end{equation}
where
\begin{subequations}\label{eq:chargeCurrentsExplicit}
\begin{align}
\hat{j}^{\mathrm{c,}0}_{\boldsymbol{r},\boldsymbol{\delta}} &= -e\frac{it}{\hbar}(\hat{c}^{\dagger}_{\boldsymbol{r}+\boldsymbol{\delta}}\hat{c}_{\boldsymbol{r}} - \hat{c}^{\dagger}_{\boldsymbol{r}}\hat{c}_{\boldsymbol{r}+\boldsymbol{\delta}}) \, , \\
\hat{j}^{\mathrm{c,a}}_{\boldsymbol{r},\pm\boldsymbol{\delta}_{\mathrm{x}}} &= \pm  e\xi_{\mathrm{R}}\frac{t}{\hbar} (\hat{c}^{\dagger}_{\boldsymbol{r}\pm\boldsymbol{\delta}_{\mathrm{x}}}\sigma_{\mathrm{y}}\hat{c}_{\boldsymbol{r}} + \hat{c}^{\dagger}_{\boldsymbol{r}}\sigma_{\mathrm{y}}\hat{c}_{\boldsymbol{r}\pm\boldsymbol{\delta}_{\mathrm{x}}}) \, , \\
\hat{j}^{\mathrm{c,a}}_{\boldsymbol{r},\pm\boldsymbol{\delta}_{\mathrm{y}}} &= \mp e\xi_{\mathrm{R}}\frac{t}{\hbar} (\hat{c}^{\dagger}_{\boldsymbol{r}\pm\boldsymbol{\delta}_{\mathrm{y}}}\sigma_{\mathrm{x}}\hat{c}_{\boldsymbol{r}} + \hat{c}^{\dagger}_{\boldsymbol{r}}\sigma_{\mathrm{x}}\hat{c}_{\boldsymbol{r}\pm\boldsymbol{\delta}_{\mathrm{y}}}) \, .
\end{align}
\end{subequations}
A finite RSOC induces the terms $\hat{j}^{\mathrm{c,a}}_{\boldsymbol{r},\boldsymbol{\delta}}$, which depend on the spin and spatial direction, as shown in Eq. \eqref{eq:chargeCurrentsExplicit}.

A spin continuity equation is derived from the time rate of change of the spin density operator $\hat{\boldsymbol{s}}_{\boldsymbol{r}}$:
\begin{align}\label{eq:spinContinuityEquationFull}
0 =& \frac{\partial \hat{\boldsymbol{s}}_{\boldsymbol{r}}}{\partial t} + \sum_{\boldsymbol{\delta}=\pm\boldsymbol{\delta}_{\mathrm{x}},\pm\boldsymbol{\delta}_{\mathrm{y}}} \hat{\boldsymbol{j}}^{\mathrm{s}}_{\boldsymbol{r},\boldsymbol{\delta}}  + J_{\mathrm{sd}} \boldsymbol{\mathcal{S}}_{\boldsymbol{r}}\times\hat{\boldsymbol{s}}_{\boldsymbol{r}} \, ,
\end{align}
where $\hat{\boldsymbol{j}}^{\mathrm{s}}_{\boldsymbol{r},\boldsymbol{\delta}}$ denotes two types of spin currents between sites $\boldsymbol{r}$ and $\boldsymbol{r}+\boldsymbol{\delta}$. In addition, the spin continuity equation (Eq. \eqref{eq:spinContinuityEquationFull}) contains the onsite torques $J_{\mathrm{sd}} \boldsymbol{\mathcal{S}}_{\boldsymbol{r}}\times\hat{\boldsymbol{s}}_{\boldsymbol{r}}$, which act on the localized spins $\boldsymbol{\mathcal{S}}_{\boldsymbol{r}}$. The two types of spin currents are denoted
\begin{equation}\label{eq:appTwoTypesSpinCurrent}
\hat{\boldsymbol{j}}^{\mathrm{s}}_{\boldsymbol{r},\boldsymbol{\delta}} = \hat{\boldsymbol{j}}^{\mathrm{s,}0}_{\boldsymbol{r},\boldsymbol{\delta}} + \hat{\boldsymbol{j}}^{\mathrm{s,a}}_{\boldsymbol{r},\boldsymbol{\delta}} \, ,
\end{equation}
where
\begin{equation}\label{eq:spinCurrentNormalHoppings}
\hat{\boldsymbol{j}}^{\mathrm{s,}0}_{\boldsymbol{r},\boldsymbol{\delta}} = \frac{it}{2}(\hat{c}^{\dagger}_{\boldsymbol{r}+\boldsymbol{\delta}}\boldsymbol{\sigma}\hat{c}_{\boldsymbol{r}} - \hat{c}^{\dagger}_{\boldsymbol{r}}\boldsymbol{\sigma}\hat{c}_{\boldsymbol{r}+\boldsymbol{\delta}}) \,
\end{equation}
is the spin current in the absence of RSOC. We label the three spin polarizations as $\hat{\boldsymbol{j}}^{\mathrm{s,}0}_{\boldsymbol{r},\boldsymbol{\delta}} = (\hat{j}^{\mathrm{s,0,x}}_{\boldsymbol{r},\boldsymbol{\delta}},\hat{j}^{\mathrm{s,0,y}}_{\boldsymbol{r},\boldsymbol{\delta}} ,\hat{j}^{\mathrm{s,0,z}}_{\boldsymbol{r},\boldsymbol{\delta}} )$. Spin is not conserved in a system with RSOC. We have not included the anomalous part $\boldsymbol{\hat{j}}^{\mathrm{s,a}}_{\boldsymbol{r},\boldsymbol{\delta}}$ in the calculation of the spin Hall conductivity. We use $\boldsymbol{\hat{j}}^{\mathrm{s,}0}_{\boldsymbol{r},\boldsymbol{\delta}}$ as defined in Eq. \eqref{eq:spinCurrentNormalHoppings} as our definition of the spin current, which we use in the Kubo formula \eqref{eq:sHallConductivity4basisGeneral}.

By labeling the three spin polarizations of the anomalous spin current as $\hat{\boldsymbol{j}}^{\mathrm{s,a}}_{\boldsymbol{r},\boldsymbol{\delta}}=(\hat{j}^{\mathrm{s,a,x}}_{\boldsymbol{r},\boldsymbol{\delta}}, \hat{j}^{\mathrm{s,a,y}}_{\boldsymbol{r},\boldsymbol{\delta}}, \hat{j}^{\mathrm{s,a,z}}_{\boldsymbol{r},\boldsymbol{\delta}})$, the exact expressions are as follows. For the $x$-component of the spin,
\begin{subequations}\label{eq:SpinCurrentsSocX}
\begin{align}
\hat{j}^{\mathrm{s,a,x}}_{\boldsymbol{r},\pm\boldsymbol{\delta}_{\mathrm{x}}}  =&  \pm\xi_{\mathrm{R}}\frac{it}{2} (\hat{c}^{\dagger}_{\boldsymbol{r}\pm\boldsymbol{\delta}_{\mathrm{x}}}\sigma_{\mathrm{z}}\hat{c}_{\boldsymbol{r}} - \hat{c}^{\dagger}_{\boldsymbol{r}}\sigma_{\mathrm{z}}\hat{c}_{\boldsymbol{r}\pm\boldsymbol{\delta}_{\mathrm{x}}} )  \, , \\
\hat{j}^{\mathrm{s,a,x}}_{\boldsymbol{r},\pm\boldsymbol{\delta}_{\mathrm{y}}}  =&  \pm\xi_{\mathrm{R}}\frac{t}{2} (\hat{c}^{\dagger}_{\boldsymbol{r}\pm\boldsymbol{\delta}_{\mathrm{y}}}\hat{c}_{\boldsymbol{r}} + \hat{c}^{\dagger}_{\boldsymbol{r}}\hat{c}_{\boldsymbol{r}\pm\boldsymbol{\delta}_{\mathrm{y}}} )  \, ,
\end{align}
\end{subequations}
for the $y$-component,
\begin{subequations}\label{eq:SpinCurrentsSocY}
\begin{align}
\hat{j}^{\mathrm{s,a,y}}_{\boldsymbol{r},\pm\boldsymbol{\delta}_{\mathrm{x}}}  =& \mp \xi_{\mathrm{R}}\frac{t}{2}(\hat{c}^{\dagger}_{\boldsymbol{r}\pm\boldsymbol{\delta}_{\mathrm{x}}}\hat{c}_{\boldsymbol{r}} + \hat{c}^{\dagger}_{\boldsymbol{r}}\hat{c}_{\boldsymbol{r}\pm\boldsymbol{\delta}_{\mathrm{x}}} )  \, , \\
\hat{j}^{\mathrm{s,a,y}}_{\boldsymbol{r},\pm\boldsymbol{\delta}_{\mathrm{y}}}  =&
\pm\xi_{\mathrm{R}}\frac{it}{2}(\hat{c}^{\dagger}_{\boldsymbol{r}\pm\boldsymbol{\delta}_{\mathrm{y}}}\sigma_{\mathrm{z}}\hat{c}_{\boldsymbol{r}} - \hat{c}^{\dagger}_{\boldsymbol{r}}\sigma_{\mathrm{z}}\hat{c}_{\boldsymbol{r}\pm\boldsymbol{\delta}_{\mathrm{y}}} )  \, ,
\end{align}
\end{subequations}
and for the $z$-component,
\begin{subequations}\label{eq:SpinCurrentsSocZ}
\begin{align}
\hat{j}^{\mathrm{s,a,z}}_{\boldsymbol{r},\pm\boldsymbol{\delta}_{\mathrm{x}}}  =&
\mp\xi_{\mathrm{R}}\frac{it}{2} (\hat{c}^{\dagger}_{\boldsymbol{r}\pm\boldsymbol{\delta}_{\mathrm{x}}}\sigma_{\mathrm{x}}\hat{c}_{\boldsymbol{r}} -  \hat{c}^{\dagger}_{\boldsymbol{r}}\sigma_{\mathrm{x}}\hat{c}_{\boldsymbol{r}\pm\boldsymbol{\delta}_{\mathrm{x}}} )  \, , \\
\hat{j}^{\mathrm{s,a,z}}_{\boldsymbol{r},\pm\boldsymbol{\delta}_{\mathrm{y}}}  =&
\mp\xi_{\mathrm{R}}\frac{it}{2} (\hat{c}^{\dagger}_{\boldsymbol{r}\pm\boldsymbol{\delta}_{\mathrm{y}}}\sigma_{\mathrm{y}}\hat{c}_{\boldsymbol{r}} - \hat{c}^{\dagger}_{\boldsymbol{r}}\sigma_{\mathrm{y}}\hat{c}_{\boldsymbol{r}\pm\boldsymbol{\delta}_{\mathrm{y}}} )  \, .
\end{align}
\end{subequations}
Note that the sign changes in $\hat{\boldsymbol{j}}^{\mathrm{s,a}}_{\boldsymbol{r},\boldsymbol{\delta}}$ when changing $\boldsymbol{\delta}\rightarrow-\boldsymbol{\delta}$, in the anomalous part of the spin current.

\section{Kubo Formula for an Antiferromagnet}\label{appendix:antiferromagnet}

In this section, we summarize the expressions for the spin Hall conductivities of an AF, together with the related single-particle eigenfunctions and eigenenergies. We focus on the solutions when the localized AF spins are out-of-plane ($\boldsymbol{n}=\pm\hat{z}$), in the $x$-$z$-plane [$\boldsymbol{n} = (\sin\theta, 0, \cos\theta )$], or in the $y$-$z$-plane [$\boldsymbol{n} = (0, \sin\theta, \cos\theta )$]. Here, $\theta$ is the polar angle relative to the $\hat{z}$-axis.

The AF eigenfunctions $\psi_{\boldsymbol{k}n}$ and eigenenergies $E_{n}(\boldsymbol{k})$ are found by diagonalizing the Hamiltonian $\hat{\mathcal{H}}$ in Eq. \eqref{eq:AFhamiltonianMomentumSpace}, written as $\hat{\mathcal{H}} = \sum_{\boldsymbol{k} \in\Diamond} \hat{\mathcal{C}}^{\dagger}_{\boldsymbol{k}}\big( \varepsilon_{0}(\boldsymbol{k}) + \mathcal{U}_{\boldsymbol{k}}\mathcal{D}_{\boldsymbol{k}}\mathcal{U}^{\dagger}_{\boldsymbol{k}}\big) \hat{\mathcal{C}}_{\boldsymbol{k}}$. The unitary matrix $\mathcal{U}_{\boldsymbol{k}}$ contains the four orthonormal eigenvectors $\psi_{\boldsymbol{k}n}$ as its column vectors. The corresponding eigenenergies are $E_{n}(\boldsymbol{k}) = \varepsilon_{0}(\boldsymbol{k}) + [\mathcal{D}_{\boldsymbol{k}}]_{nn}$, where $\varepsilon_{0}(\boldsymbol{k})$ is the hopping energy [Eq. \eqref{eq:TBhoppingEnergyMomentumSpace}] and $\mathcal{D}_{\boldsymbol{k}}$ is a diagonal matrix. The spin Hall conductivities are calculated by using Eq. \eqref{eq:sHallConductivity4basisGeneral} together with $\psi_{\boldsymbol{k}n}$ and $E_{n}(\boldsymbol{k})$.

In the case of out-of-plane AF spins ($\boldsymbol{n}=\pm\hat{z}$), the eigenfunctions, eigenenergies and spin Hall conductivity are found by considering the limits $\cos\theta\rightarrow\pm1$ in the corresponding expressions obtained for AF spins in the $x$-$z$-plane or the $y$-$z$-plane, as summarized in Apps. \ref{appendix:xzPlaneSolutions} and \ref{appendix:yzPlaneSolutions}, respectively. When $\boldsymbol{n}=\hat{z}$, the eigenenergies are doubly degenerate, as $\mathcal{D}_{\boldsymbol{k}}=\mathrm{diag}(\Delta^{\mathrm{z}}_{-}, \Delta^{\mathrm{z}}_{+},\Delta^{\mathrm{z}}_{-}, \Delta^{\mathrm{z}}_{+})$, with $\Delta^{\mathrm{z}}_{\pm}$ from Eq. \eqref{eq:twoAFeigenEnergiesOutOfPLane}. For out-of-plane AF spins, the only nonzero terms in Eq. \eqref{eq:sHallConductivity4basisGeneral} stem from the expectation values between states with different energies, which yields the finite spin Hall conductivity $\sigma^{\mathrm{z}}_{\mathrm{sH}}$ in Eq. \eqref{eq:spinHconductOutPlaneSpinsExprs}.

For brevity of notation, we define the parameters
\begin{subequations}
\begin{align}
\mathcal{M} =& -t\xi_{\mathrm{sd}} \, , \\
\mathcal{R} =& 2t\xi_{\mathrm{R}} \, , \\
\mathcal{X} =& \sin k_{\mathrm{x}}a \, , \\
\mathcal{Y} =& \sin k_{\mathrm{y}}a \, ,
\end{align}
\end{subequations}
which we use in the following.

\subsection{Neel Order Parameter in the $x$-$z$-plane}\label{appendix:xzPlaneSolutions}

Localized AF spins in the $x$-$z$-plane are described by the Neel order parameter $\boldsymbol{n} = (\sin\theta, 0, \cos\theta )$. We diagonalize the Hamiltonian $\hat{\mathcal{H}}$ in Eq. \eqref{eq:AFhamiltonianMomentumSpace}. The relevant terms are labeled $\hat{\mathcal{H}} = \sum_{\boldsymbol{k} \in\Diamond} \hat{\mathcal{C}}^{\dagger}_{\boldsymbol{k}}\big( \varepsilon_{0}(\boldsymbol{k}) + \mathcal{U}^{\mathrm{xz}}_{\boldsymbol{k}}\mathcal{D}^{\mathrm{xz}}_{\boldsymbol{k}}\mathcal{U}^{\mathrm{xz}\dagger}_{\boldsymbol{k}}\big) \hat{\mathcal{C}}_{\boldsymbol{k}}$.

The eigenenergies consist of $\varepsilon_{0}(\boldsymbol{k})$ plus one of the eigenvalues determined by $\mathcal{D}^{\mathrm{xz}}_{\boldsymbol{k}} = \mathrm{diag}(\Delta^{\mathrm{xz}}_{1,-},\Delta^{\mathrm{xz}}_{1,+},\Delta^{\mathrm{xz}}_{2,-},\Delta^{\mathrm{xz}}_{2,+})$, written explicitly as
\begin{subequations}\label{eq:eigenvaluesXZ}
\begin{align}
\frac{ \Delta^{\mathrm{xz}}_{1,\pm} }{t} =& \pm \sqrt{ \xi^{2}_{\mathrm{sd}} + 4\xi^{2}_{\mathrm{R}} (\mathcal{X}^{2} + \mathcal{Y}^{2} ) +4\xi_{\mathrm{sd}}\xi_{\mathrm{R}} \mathcal{Y} \sin\theta  }    \, , \\
\frac{ \Delta^{\mathrm{xz}}_{2,\pm} }{t} =& \pm \sqrt{ \xi^{2}_{\mathrm{sd}} + 4\xi^{2}_{\mathrm{R}}(\mathcal{X}^{2} + \mathcal{Y}^{2} ) - 4\xi_{\mathrm{sd}}\xi_{\mathrm{R}}\mathcal{Y} \sin\theta  }    \, .
\end{align}
\end{subequations}
Corresponding to the four eigenvalues $\Delta^{\mathrm{xz}}_{1,\pm}$ and $\Delta^{\mathrm{xz}}_{2,\pm}$, the eigenfunctions are $\psi^{\mathrm{xz}}_{1,\pm}$ and $\psi^{\mathrm{xz}}_{2,\pm}$. The unitary matrix $\mathcal{U}^{\mathrm{xz}}$ is written in block form as $\mathcal{U}^{\mathrm{xz}} = \begin{pmatrix} \psi^{\mathrm{xz}}_{1,-}, \psi^{\mathrm{xz}}_{1,+}, \psi^{\mathrm{xz}}_{2,-}, \psi^{\mathrm{xz}}_{2,+}  \end{pmatrix}$. The eigenfunctions have four components, written as
\begin{subequations}\label{eq:psiXZ}
\begin{align}
\psi^{\mathrm{xz}}_{1,\pm} =&  \frac{  1 }{ 2\sqrt{ \Delta^{\mathrm{xz}}_{1,\pm}}  \sqrt{ \Delta^{\mathrm{xz}}_{1,\pm} + \mathcal{M} \cos\theta } }   \begin{pmatrix} \gamma_{1,\pm} \\ -\gamma_{1,\pm} \end{pmatrix}    \, , \\
\psi^{\mathrm{xz}}_{2,\pm} =& \frac{  1 }{ 2\sqrt{ \Delta^{\mathrm{xz}}_{2,\pm}}  \sqrt{ \Delta^{\mathrm{xz}}_{2,\pm} - \mathcal{M} \cos\theta } } \begin{pmatrix} \gamma_{2,\pm} \\   \gamma_{2,\pm} \end{pmatrix}     \, ,
\end{align}
\end{subequations}
in terms of
\begin{subequations}\label{eq:gammas}
\begin{align}
\gamma_{1,\pm} =&  \begin{pmatrix}  \mathcal{M}\sin\theta - \mathcal{R}(\mathcal{Y} + i\mathcal{X})   \\  \mathcal{M}\cos\theta  -  \Delta^{\mathrm{xz}}_{1,\pm}  \end{pmatrix}    \, , \\
\gamma_{2,\pm} =&   \begin{pmatrix}  \mathcal{M}\sin\theta + \mathcal{R}(\mathcal{Y} + i\mathcal{X})  \\  \mathcal{M}\cos\theta + \Delta^{\mathrm{xz}}_{2,\pm} \end{pmatrix}     \, .
\end{align}
\end{subequations}
The spin Hall conductivity is calculated from Eq. \eqref{eq:sHallConductivity4basisGeneral} together with Eqs. \eqref{eq:eigenvaluesXZ}, \eqref{eq:psiXZ} and \eqref{eq:gammas}. For the $z$-polarization of the spin, the spin Hall conductivity $\sigma^{\mathrm{z,xz}}_{\mathrm{sH}}$ when the AF spins rotate in the $x$-$z$-plane is
\begin{align}\label{eq:appXZplanSigmaZgenereltUttrykk}
\sigma^{\mathrm{z,xz}}_{\mathrm{sH}} =&  \frac{(2\pi)^{2}}{N^{2} /2 } \sum_{\boldsymbol{k}\in\Diamond}  \frac{\cos k_{\mathrm{x}}a}{ 2\pi} \Bigg[  \Big(   \frac{ f^{\mathrm{xz}}_{1,-}-f^{\mathrm{xz}}_{1,+} }{ 2( \frac{\Delta^{\mathrm{xz}}_{1,+} }{ t})^{3} }  -  \frac{ f^{\mathrm{xz}}_{2,-}-f^{\mathrm{xz}}_{2,+} }{ 2 (\frac{\Delta^{\mathrm{xz}}_{2,+} }{ t})^{3} }   \Big)    \nonumber \\
& \times 2\xi_{\mathrm{R}}\xi_{\mathrm{sd}} \mathcal{Y} \sin\theta \nonumber \\
& +\Big(    \frac{ f^{\mathrm{xz}}_{1,-}-f^{\mathrm{xz}}_{1,+} }{ 2( \frac{\Delta^{\mathrm{xz}}_{1,+} }{ t})^{3} }  +  \frac{ f^{\mathrm{xz}}_{2,-}-f^{\mathrm{xz}}_{2,+} }{ 2 (\frac{\Delta^{\mathrm{xz}}_{2,+} }{ t})^{3} }   \Big)(2\xi_{\mathrm{R}})^{2} \mathcal{Y}^{2}   \Bigg] \, ,
\end{align}
where $f^{\mathrm{xz}}_{1,\pm} = f_{\mathrm{FD}}(\varepsilon_{0}+\Delta^{\mathrm{xz}}_{1,\pm} - \mu)$, and similarly for $f^{\mathrm{xz}}_{2,\pm}$.

For the $x$-polarization of the spin, the spin Hall conductivity is
\begin{align}\label{eq:appXZplanSigmaX}
\sigma^{\mathrm{x,xz}}_{\mathrm{sH}} =&   \frac{(2\pi)^{2}}{N^{2} /2 } \sum_{\boldsymbol{k}\in\Diamond}  \frac{\cos k_{\mathrm{x}}a}{ 2\pi}  \Big(   \frac{ f^{\mathrm{xz}}_{1,-}-f^{\mathrm{xz}}_{1,+} }{ 2( \frac{\Delta^{\mathrm{xz}}_{1,+} }{ t})^{3} }  -  \frac{ f^{\mathrm{xz}}_{2,-}-f^{\mathrm{xz}}_{2,+} }{ 2 (\frac{\Delta^{\mathrm{xz}}_{2,+} }{ t})^{3} }   \Big) \nonumber \\
& \times 2\xi_{\mathrm{R}}\xi_{\mathrm{sd}} \mathcal{Y}  \cos\theta  \, .
\end{align}

The integrands in the expressions of $\sigma^{\mathrm{z,xz}}_{\mathrm{sH}}$ and $\sigma^{\mathrm{x,xz}}_{\mathrm{sH}}$, in Eqs. \eqref{eq:appXZplanSigmaZgenereltUttrykk} and \eqref{eq:appXZplanSigmaX}, respectively, are both even in $k_{\mathrm{x}}$ and even in $k_{\mathrm{y}}$.

For the $y$-polarization of the spin, the spin Hall conductivity $\sigma^{\mathrm{y,xz}}_{\mathrm{sH}}$ vanishes.

\subsection{Neel Order Parameter in the $y$-$z$-plane}\label{appendix:yzPlaneSolutions}

In the case where the AF spins rotate in the $y$-$z$-plane, the Neel order parameter is $\boldsymbol{n} = (0, \sin\theta, \cos\theta )$. Here, the Hamiltonian $\hat{\mathcal{H}} = \sum_{\boldsymbol{k} \in\Diamond} \hat{\mathcal{C}}^{\dagger}_{\boldsymbol{k}}\big( \varepsilon_{0}(\boldsymbol{k}) + \mathcal{U}^{\mathrm{yz}}_{\boldsymbol{k}}\mathcal{D}^{\mathrm{yz}}_{\boldsymbol{k}}\mathcal{U}^{\mathrm{yz}\dagger}_{\boldsymbol{k}}\big) \hat{\mathcal{C}}_{\boldsymbol{k}}$ is diagonalized by $\mathcal{D}^{\mathrm{yz}}_{\boldsymbol{k}} = \mathrm{diag}(\Delta^{\mathrm{yz}}_{1,-},\Delta^{\mathrm{yz}}_{1,+},\Delta^{\mathrm{yz}}_{2,-},\Delta^{\mathrm{yz}}_{2,+})$, with the four eigenvalues
\begin{subequations}\label{eq:eigenvaluesYZ}
\begin{align}
\frac{ \Delta^{\mathrm{yz}}_{1,\pm} }{t} =& \pm \sqrt{ \xi^{2}_{\mathrm{sd}} + 4\xi^{2}_{\mathrm{R}} (\mathcal{X}^{2} + \mathcal{Y}^{2} ) +4\xi_{\mathrm{sd}}\xi_{\mathrm{R}} \mathcal{X} \sin\theta  }    \, , \\
\frac{ \Delta^{\mathrm{yz}}_{2,\pm} }{t} =& \pm \sqrt{ \xi^{2}_{\mathrm{sd}} + 4\xi^{2}_{\mathrm{R}} (\mathcal{X}^{2} + \mathcal{Y}^{2} ) - 4\xi_{\mathrm{sd}}\xi_{\mathrm{R}} \mathcal{X} \sin\theta  }    \, .
\end{align}
\end{subequations}
The unitary matrix is $\mathcal{U}^{\mathrm{yz}} = \begin{pmatrix} \psi^{\mathrm{yz}}_{1,-}, \psi^{\mathrm{yz}}_{1,+}, \psi^{\mathrm{yz}}_{2,-}, \psi^{\mathrm{yz}}_{2,+}  \end{pmatrix}$, written in terms of the eigenfunctions
\begin{subequations}\label{eq:psiyZ}
\begin{align}
\psi^{\mathrm{yz}}_{1,\pm} =&  \frac{  1 }{ 2\sqrt{ \Delta^{\mathrm{yz}}_{1,\pm}}  \sqrt{ \Delta^{\mathrm{yz}}_{1,\pm} - \mathcal{M} \cos\theta } }  \begin{pmatrix} \kappa_{1,\pm} \\ \kappa_{1,\pm} \end{pmatrix}    \, , \\
\psi^{\mathrm{yz}}_{2,\pm} =&  \frac{  1 }{ 2\sqrt{ \Delta^{\mathrm{yz}}_{2,\pm}}  \sqrt{ \Delta^{\mathrm{yz}}_{2,\pm} + \mathcal{M} \cos\theta } }  \begin{pmatrix} \kappa_{2,\pm} \\   -\kappa_{2,\pm} \end{pmatrix}     \, ,
\end{align}
\end{subequations}
where
\begin{subequations}\label{eq:kappas}
\begin{align}
\kappa_{1,\pm} =&   \begin{pmatrix} -i  \mathcal{M}\sin\theta + \mathcal{R}(\mathcal{Y} + i\mathcal{X})   \\   \Delta^{\mathrm{yz}}_{1,\pm} -  \mathcal{M}\cos\theta    \end{pmatrix}    \, , \\
\kappa_{2,\pm} =&  \begin{pmatrix} -i  \mathcal{M}\sin\theta -\mathcal{R}(\mathcal{Y} + i\mathcal{X})  \\  - \Delta^{\mathrm{yz}}_{2,\pm} - \mathcal{M}\cos\theta   \end{pmatrix}     \, .
\end{align}
\end{subequations}
Based on Eqs. \eqref{eq:sHallConductivity4basisGeneral}, \eqref{eq:eigenvaluesYZ}, \eqref{eq:psiyZ} and \eqref{eq:kappas}, the spin Hall conductivity for AF spins in the $y$-$z$-plane yields
\begin{align}\label{eq:sHallzpolariasjonYZplan}
\sigma^{\mathrm{z,yz}}_{\mathrm{sH}} =&  \frac{(2\pi)^{2}}{N^{2} /2 } \sum_{\boldsymbol{k}\in\Diamond} \frac{ \cos k_{\mathrm{x}}a }{ 2\pi} \Big(   \frac{f^{\mathrm{yz}}_{1,-}-f^{\mathrm{yz}}_{1,+} }{2(\frac{\Delta^{\mathrm{yz}}_{1,+} }{ t})^{3}  }  + \frac{f^{\mathrm{yz}}_{2,-}-f^{\mathrm{yz}}_{2,+} }{2(\frac{\Delta^{\mathrm{yz}}_{2,+} }{ t})^{3} }   \Big) \nonumber \\
&\times(2\xi_{\mathrm{R}})^{2} \mathcal{Y}^{2}  \,
\end{align}
when we consider the $z$-component of the spin. In Eq. \eqref{eq:sHallzpolariasjonYZplan}, $f^{\mathrm{yz}}_{1,\pm} = f_{\mathrm{FD}}(\varepsilon_{0}+\Delta^{\mathrm{yz}}_{1,\pm} - \mu)$, and similarly for $f^{\mathrm{yz}}_{2,\pm}$. The integrand in Eq. \eqref{eq:sHallzpolariasjonYZplan} is even in $k_{\mathrm{x}}$ and even in $k_{\mathrm{y}}$.

For the $x$-polarization of the spin, the expression for the spin Hall conductivity is
\begin{align}\label{eq:sHallxpolariasjonYZplan}
\sigma^{\mathrm{x,yz}}_{\mathrm{sH}} =&  \frac{(2\pi)^{2}}{N^{2} /2 } \sum_{\boldsymbol{k}\in\Diamond} \frac{\cos k_{\mathrm{x}} a }{2\pi }  \Big[  \frac{f^{\mathrm{yz}}_{1,-}-f^{\mathrm{yz}}_{1,+} }{2(\frac{\Delta^{\mathrm{yz}}_{1,+} }{ t})^{3}  }  - \frac{f^{\mathrm{yz}}_{2,-}-f^{\mathrm{yz}}_{2,+} }{2(\frac{\Delta^{\mathrm{yz}}_{2,+} }{ t})^{3} }  \Big] \nonumber \\
& \times2 \xi_{\mathrm{R}} \xi_{\mathrm{sd}} \mathcal{Y} \cos \theta \nonumber \\
=& 0 \, ,
\end{align}
where $\sigma^{\mathrm{x,yz}}_{\mathrm{sH}}$ vanishes because the integrand in Eq. \eqref{eq:sHallxpolariasjonYZplan} is antisymmetric in both $k_{\mathrm{x}}$ and $k_{\mathrm{y}}$.

The spin Hall conductivity $\sigma^{\mathrm{y,yz}}_{\mathrm{sH}}$ for the $y$-polarization of the spin vanishes because the relevant expectation values in Eq. \eqref{eq:sHallConductivity4basisGeneral} have no finite imaginary part.


\end{document}